%% file: subgiants.tex
\PassOptionsToPackage{unicode}{hyperref}
\PassOptionsToPackage{hyphens}{url}
\PassOptionsToPackage{dvipsnames,svgnames*,x11names*}{xcolor}
\documentclass[
  astrosymb, twocolumn, tighten]{aastex63}
\usepackage{amsmath,amssymb}
\usepackage{lmodern}
\usepackage{ifxetex,ifluatex}
\ifnum 0\ifxetex 1\fi\ifluatex 1\fi=0 
  \usepackage[T1]{fontenc}
  \usepackage[utf8]{inputenc}
  \usepackage{textcomp} 
\else 
  \usepackage{unicode-math}
  \defaultfontfeatures{Scale=MatchLowercase}
  \defaultfontfeatures[\rmfamily]{Ligatures=TeX,Scale=1}
\fi
\IfFileExists{upquote.sty}{\usepackage{upquote}}{}
\IfFileExists{microtype.sty}{
  \usepackage[]{microtype}
  \UseMicrotypeSet[protrusion]{basicmath} 
}{}
\makeatletter
\@ifundefined{KOMAClassName}{
  \IfFileExists{parskip.sty}{%
    \usepackage{parskip}
  }{
    \setlength{\parindent}{0pt}
    \setlength{\parskip}{6pt plus 2pt minus 1pt}}
}{
  \KOMAoptions{parskip=half}}
\makeatother
\usepackage{xcolor}
\IfFileExists{xurl.sty}{\usepackage{xurl}}{} 
\IfFileExists{bookmark.sty}{\usepackage{bookmark}}{\usepackage{hyperref}}
\hypersetup{
  colorlinks=true,
  linkcolor=Maroon,
  filecolor=Maroon,
  citecolor=Blue,
  urlcolor=Blue,
  pdfcreator={LaTeX via pandoc}}
\usepackage{longtable,booktabs,array}
\usepackage{calc} 
\usepackage{etoolbox}
\makeatletter
\patchcmd\longtable{\par}{\if@noskipsec\mbox{}\fi\par}{}{}
\makeatother
\IfFileExists{footnotehyper.sty}{\usepackage{footnotehyper}}{\usepackage{footnote}}
\makesavenoteenv{longtable}
\usepackage{graphicx}
\makeatletter
\def\maxwidth{\ifdim\Gin@nat@width>\linewidth\linewidth\else\Gin@nat@width\fi}
\def\maxheight{\ifdim\Gin@nat@height>\textheight\textheight\else\Gin@nat@height\fi}
\makeatother
\setkeys{Gin}{width=\maxwidth,height=\maxheight,keepaspectratio}
\makeatletter
\def\fps@figure{htbp}
\makeatother
\providecommand{\tightlist}{%
  \setlength{\itemsep}{0pt}\setlength{\parskip}{0pt}}
\input{macros.tex}
\usepackage{mathptmx,txfonts,tikz}
\ifluatex
  \usepackage{selnolig}  
\fi
\usepackage[]{natbib}
\date{\today}

\begin{document}

\shorttitle{Mixed Modes and Surface Effects II}
\title{Mixed Modes and Asteroseismic Surface Effects: II. Subgiant Systematics}
\input{preamble}
\begin{abstract}
Models of solar-like oscillators yield acoustic modes at different frequencies than would be seen in actual stars possessing identical interior structure, due to modelling error near the surface. This asteroseismic ``surface term'' must be corrected when mode frequencies are used to infer stellar structure. Subgiants exhibit oscillations of mixed acoustic ($p$-mode) and gravity ($g$-mode) character, which defy description by the traditional $p$-mode asymptotic relation. Since nonparametric diagnostics of the surface term rely on this description, they cannot be applied to subgiants directly. In Paper I, we generalised such nonparametric methods to mixed modes, and showed that traditional surface-term corrections only account for mixed-mode coupling to, at best, first order in a perturbative expansion. Here, we apply those results, modelling subgiants using asteroseismic data. We demonstrate that, for grid-based inference of subgiant properties using individual mode frequencies, neglecting higher-order effects of mode coupling in the surface term results in significant systematic differences in the inferred stellar masses, and measurable systematics in other fundamental properties. While these systematics are smaller than those resulting from other choices of model construction, they persist for both parametric and nonparametric formulations of the surface term. This suggests that mode coupling should be fully accounted for when correcting for the surface term in seismic modelling with mixed modes, irrespective of the choice of correction used. The inferred properties of subgiants, in particular masses and ages, also depend on the choice of surface-term correction, in a different manner from both main-sequence and red giant stars.
\end{abstract}
\defcitealias{ball_correction_2014}{BG14}
\defcitealias{grevesse_standard_1998}{GS98}
\defcitealias{ong_semianalytic_2020}{OB20}
\newcommand\bg{{\citetalias{ball_correction_2014}}}
\newcommand\gs{{\citetalias{grevesse_standard_1998}}}
\newcommand\ob{{\citetalias{ong_semianalytic_2020}}}
\newcommand\mesa{{\textsc{mesa}}}
\newcommand\gyre{{\textsc{gyre}}}
\newcommand\diamonds{{\textsc{diamonds}}}

\hypertarget{introduction}{%
\section{Introduction}\label{introduction}}

Solar-like oscillators exhibit a characteristic comb-like pattern of
peaks in the power spectra of their photometric or velocimetric
time-series, where each peak occurs at the natural frequency associated
with a normal mode of oscillation. For pressure waves (p-modes) in
Sun-like stars, the frequencies of these normal modes satisfy an
approximate asymptotic relation
\citep[e.g.][]{tassoul_second-order_1994} \begin{equation}
\nu_{n,l,m} \sim \Dnu\left(n + {l \over 2} + \epsilon_{l,n}\right),\label{eq:asymptotic}
\end{equation} where \(n\) is the radial order of a mode, \(l\) is its
angular degree, \(\epsilon_{l,n}\) is a phase offset, and \Dnu, the
large frequency separation, is the characteristic frequency of the
p-mode cavity. These frequencies depend on only the interior structure
of these stars, and forward modelling is needed in order to constrain
other stellar properties and physics, such as their ages and mixing
parameters, which best reproduce this structure. On the other hand,
detailed examination of the solar case reveals that the normal modes of
standard solar models, with interior structures matching that of the
Sun, have frequencies differing from the corresponding observed
normal-mode oscillation frequencies. These frequency differences are
thought to derive primarily from modelling errors localised to the solar
surface \citep[e.g.][]{jcd_theory_1991, rosenthal_convective_1999}, and
therefore are collectively referred to as the solar ``surface term''. It
is believed that such a surface term should also exist in seismically
constrained models of other stars as well.

These frequency differences must be accounted for when modelling stars
to match observed mode frequencies, and many prescriptions exist to
correct for or diagnose their effects. These include both parametric
``surface term corrections,'' which describe the surface term as a
slowly-varying function of frequency
\citep[e.g.][]{kjeldsen_correcting_2008, ball_correction_2014, sonoi_surface_2015},
as well as nonparametric techniques, which exploit the structure of the
asymptotic relation \cref{eq:asymptotic}
\citep{otifloranes_use_2005, roxburgh_ratio_2005, roxburgh_asteroseismic_2016}.
\citet{ong_differential_2021} demonstrated that the inference of some
stellar properties, like stellar masses, radii, and ages, is robust to
variations in choice of method on the main sequence, but exhibits
nontrivial systematic dependence on surface-term methodology for a
sample of red giants. They suggested that the difference between the two
regimes might be illuminated by examining stars in an intermediate stage
of evolution --- subgiants.

Subgiants are of particular interest in the asteroseismic context.
Generally speaking, nonradial modes can be classified into p-modes
propagating in the exterior of a star, and internal gravity waves
(g-modes) propagating in its radiative interior
\citep{unno_nonradial_1989}. The evolution of stars off the main
sequence causes the maximum Brunt-Väisälä frequency of the interior
g-mode cavity to increase rapidly as the star expands, and its radiative
core contracts, in the process of crossing the Hertzsprung gap. As these
g-modes approach the frequencies of the p-modes, they couple to produce
modes of mixed character, resulting in a forest of avoided crossings
\citep{aizenman_avoided_1977}. The specific configurations of these
``mixed modes'' strongly constrain the interior structure of the star,
especially near the evanescent region between the two mode cavities
\citep[e.g.][]{deheuvels_constraints_2011, noll_probing_2021}.

Subgiants are also astrophysically interesting in a broader sense,
insofar as they may shed light on the physics underlying
post-main-sequence evolution \citep[cf.][ and references
therein]{godoyrivera_testing_2021}, the evolution of planetary systems
which they may host \citep[e.g.][]{huber_saturn_2019}, or the histories
of the stellar populations of which they are members
\citep[e.g.][]{chaplin_age_2020}. Historically, they have been less
well-studied than both main-sequence stars and red giants, as a result
of unfavourable observational selection functions: they are both much
less common than main-sequence stars, and much fainter than red giants.
However, for detailed asteroseismic characterisation, these
considerations are in some ways reversed: they are at once more
observationally accessible than main-sequence stars (in consequence of
larger photometric amplitudes and longer dynamical timescales), and more
computationally tractable than red giants
\citep[cf.][]{rg_1_2020, rg_2_2020}. For this reason, subgiants
constitute a nontrivial portion of the asteroseismic targets for the
TESS \citep{schofield_atl_2019} mission. The upcoming PLATO mission
\citep{rauer_plato_2014} is expected to yield even more subgiant targets
amenable to seismic analysis.

However, mixed modes in subgiants raise methodological issues, in that
they cease to satisfy \cref{eq:asymptotic}. Since the theoretical
constructions underlying nonparametric characterisations of the surface
term rely on \cref{eq:asymptotic}, which strictly applies only to
p-modes, these techniques are applicable only to the extent that
\cref{eq:asymptotic} is valid; in their unmodified form they are
therefore inapplicable to mixed modes. Additionally, existing parametric
surface-term corrections have all been developed for p-modes, and the
theoretical basis for applying them to mixed modes has hitherto been
less certain. These methodological concerns must be addressed if
detailed seismic characterisation of subgiants, in spite of the surface
term, is to be credible.

While mixed modes have been observed in both subgiants and red giants,
such mode mixing has so far been much less of an issue in red giants.
Traditionally, descriptions of the coupling between the two mode
cavities have relied on the Jeffreys-Wentzel-Kramers-Brillouin (JWKB)
approximation \citep{takata_asymptotic_2016, pincon_probing_2020}, which
describes the interior g-mode cavities in evolved red giants extremely
well. In this regime, the density of mixed modes is sufficiently high,
and the coupling between the two mode cavities is also sufficiently
weak, that the modes with the highest local amplitudes behave
essentially like p-modes \citep[e.g.][]{aertsbook}. For quadrupole and
higher-degree modes, this permits mode mixing to be essentially ignored,
and nonparametric methods to be applied to the most p-dominated mixed
modes as if they were pure p-modes, with little loss of accuracy
\citep{ball_surface_2018, jorgensen_investigating_2020, ong_differential_2021}.
Unfortunately, none of these conditions hold in subgiants. Consequently,
the generalisation of nonparametric treatments of the surface term to
mixed modes in the strongly nonasymptotic regime, which is necessary for
analysis of subgiants and for dipole mixed modes in red giants, has been
a longstanding open theoretical and methodological problem.

In the companion paper to this work \citep[hereafter Paper
I]{ong_surface_1}, we derived a generalisation for one class of
nonparametric treatments of the surface term --- that of
\(\epsilon_l\)-matching, in the sense of
\citet{roxburgh_asteroseismic_2016}. Assuming that a nonparametric
diagnostic for the surface term of this kind can be determined for the
pure p-modes, we prescribed an analogous likelihood function for mixed
modes, which quantifies whether a set of differences between observed
and model mixed modes is consistent with both a surface-localised
structural perturbation, as well as with the eigenvalue equation of the
internal coupling matrices associated with the stellar model
\citep[hereafter \ob]{ong_semianalytic_2020}. In the process of doing
so, we also derived generalisations of some existing parametric
descriptions --- in particular that of \citet{ball_correction_2014}
(hereafter \bg) --- to fully account for mixed-mode coupling, within the
coupling-matrix construction. This algebraic approach does not rely on
the JWKB approximation, and therefore is suitable for application to
subgiants, unlike existing techniques.

In this work, we investigate the viability of these generalised
prescriptions in the inverse direction --- applying them in inferring
the global and structural properties of subgiants. We will examine
systematic differences in the inferred values of global parameters ---
masses, radii, ages, and initial helium abundances --- as returned from
different treatments of the surface term, all else being equal. In this
manner, we fill in the existing gap between previous characterisations
of surface-related modelling systematics on the main sequence
\citep{basu_robustness_2018, nsamba_asteroseismic_2018, compton_surface_2018}
and in red giants
\citep{jorgensen_investigating_2020, ong_differential_2021}. We describe
our modelling procedure in \autoref{sec:modelling}, and our subgiant
sample in \autoref{sec:sample}. In \autoref{sec:results}, we perform
statistical tests to ascertain the relative sizes of various systematic
effects; in \autoref{sec:discussion} we discuss the implications that
these results may have for the use of individual mode frequencies in
stellar modelling, and make recommendations for best practices going
forward.

\hypertarget{modelling-procedure}{%
\section{Modelling Procedure}\label{modelling-procedure}}

\label{sec:modelling}

While there exists significant methodological variability in how such
seismic modelling is to be done, we can, broadly speaking, make a
distinction between large-scale multi-target grid searches \citep[as in
e.g.][]{mckeever_helium_2019, jorgensen_investigating_2020, li_ages_2020, nsamba_asteroseismic_2020, ong_differential_2021}
and more ``boutique'' target-by-target modelling \citep[as in
e.g.][]{ball_surface_2018, huber_saturn_2019, chaplin_age_2020, ball_robust_2020, noll_probing_2021}.
In the former case, a precomputed grid of stellar models spanning a
reasonably large parameter space is used, and the goal of the modelling
exercise is to search for the model within the grid that best matches a
set of observational constraints, for each target in the sample.
Conversely, in the latter case, the region of parameter space being
searched is adapted to each target, and hence is much smaller; the
stellar models used in the procedure are also typically constructed ad
hoc for each target. These two approaches are not necessarily mutually
exclusive. The results of grid searches may be used to restrict the
region of parameter space being considered for more refined analysis.
While boutique modelling often involves an optimisation-based parameter
search, it is also not uncommon to construct more finely-sampled
``detailed'' grids adapted to individual targets. For this study,
however, we restrict our attention to the former case, using a single
precomputed model grid for all stars.

\hypertarget{model-grids}{%
\subsection{Model grids}\label{model-grids}}

We construct a grid of subgiant models with
\mesa~\citep{mesa_paper_1, mesa_paper_2, mesa_paper_4} version 12778,
using Eddington-grey atmospheres and the solar elemental mixture of
\citet{grevesse_standard_1998}. The parameters of the grid were the
stellar mass (in the range \(M \in [0.8 M_\odot, 2.0 M_\odot]\)),
initial helium abundance (\(Y_0 \in [0.248, 0.32]\)), initial
metallicity (\(\feh_0 \in [-1, 0.5]\)), the mixing length parameter
(\(\amlt \in [1.3, 2.2]\)), and the core step overshoot parameter
\(\fov\), whose distribution we describe in more detail below. Diffusion
and settling of helium and heavy elements was handled using the
formulation of \citet{thoul_element_1994}, with an additional
mass-dependent prefactor to the diffusion coefficient \citep[as
prescribed in][]{viani_investigating_2018} to smoothly disable diffusion
at higher masses. For each evolutionary track, we retained stellar
models starting when \(\numax\Delta\Pi_1 = 2\) (i.e.~where the local
spacing of dipole \(g\)-modes is comparable to \numax), and ending when
\(\Dnu=9\ \mu\)Hz.

The first four of these grid parameters were distributed uniformly using
joint Sobol sequences of length \(2^{13}-1 = 8191\) over the parameter
space described above. For the overshoot parameters, we opted to use the
mass-overshoot relation of \citet{viani_overshoot_2020}, rather than
specifying values of \fov~uniformly using Sobol sequences. In order to
do so, we needed to translate between the different implementations of
step overshooting in YREC \citep{demarque_yrec_2008} --- the stellar
evolution code used in \citet{viani_overshoot_2020} --- and \mesa, as
well as between different definitions of how the input values of the
overshoot parameter are related to the final, effective overshooting
distance. In particular, the main results of \cite{viani_overshoot_2020}
concern YREC's implementation of instantaneous overshoot mixing, in
which the convective boundary for a convective core of size
\(R_\text{cz}\) is extended outwards by a distance \begin{equation}
    L_{\text{ov},\text{YREC}} = {\fov \over {1 \over H_p} + {1 \over R_\text{cz}}},\label{eq:yreceff}
\end{equation} by directly modifying the superadiabatic gradient
\(\nabla - \nabla_\text{ad}\) before solving the structure equations;
here \(H_p\) is the pressure scale height at the convective boundary.
Note that this ``effective'' overshooting distance is different from the
nominal ``input'' overshooting length \(\fov H_p\); the mass-overshoot
relation of \cite{viani_overshoot_2020} yields values of
\(f_\text{eff} = L_\text{ov}/H_p\) as a function of stellar mass, rather
than \fov~per se. In contrast, \mesa~implements diffusive overshoot
mixing (or ``overmixing'') following the prescription of
\citet{herwig_evolution_2000}, where the effective diffusion coefficient
\(D\) for the mixing-length-theory treatment of convection, which would
otherwise vanish outside the convective core, is artificially increased.
In \mesa's ``step overshoot'' mode, this is set to a uniform value
\(D_0 = D(R_\text{cz} - f_0 H_p)\) for a region extending
\begin{equation}
    L_{\text{ov},\text{\mesa}} = \fov \cdot \max(R_\text{cz}/\alpha_\text{ov}, H_p) \label{eq:mesaeff}
\end{equation} outwards from the convective core boundary, where
\(\alpha_\text{ov}\) is a free parameter which we set to 1 for our
computations.

\begin{figure}
\centering
\includegraphics{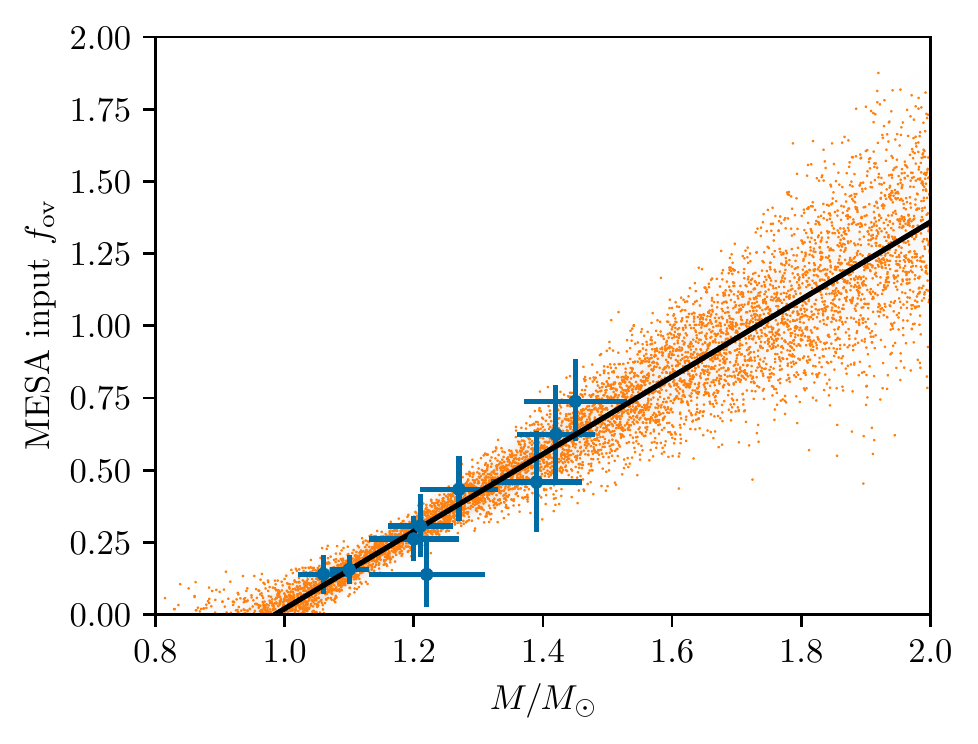}
\caption{Input core overshoot values used in the model grid. The sample
of \citet{viani_overshoot_2020} is shown with the blue points, with a
fitted line shown in black. We distribute values of \(\fov\) within the
\(2\sigma\) region around the fitted curve (orange
points).\label{fig:ov}}
\end{figure}

To reconcile these differences, for each star in the sample of
\citet{viani_overshoot_2020}, we constructed \mesa~models with identical
global properties and input physics as their corresponding YREC models,
for a range of input values of \(\fov\). For each of these models we
found \(L_\text{ov}\), and therefore \(f_\text{eff}\), directly by
comparison with a fiducial model that was constructed with \(\fov=0\).
We then solved for the values of \(\fov\) that yielded the same values
of \(f_\text{eff}\) as the YREC models used in
\citet{viani_overshoot_2020}. We show the results of doing this in
\cref{fig:ov}. From this, we fitted a mass-overshoot relation with
respect to the input values of \fov~(black line). For each point in this
grid, we then sampled \(\fov\) randomly from the \(\pm 2\sigma\) region
around this fitted mass-overshoot relation (shown with orange points in
\cref{fig:ov}), treating negative values as 0.

For each model in the grid \(i\) and star \(j\), we construct a cost
function \begin{equation}
    \chi^2_{i,j} = \left(\feh_i - \feh_j \over e_{\feh,j}\right)^2 + \left(\teff_i - \teff_j \over e_{\teff,j}\right)^2 + \chi^2_{\text{seis},j},\label{eq:chi2}
\end{equation} where the final term is a seismic cost function
determined by the choice of surface term correction under consideration.
Note that while \Dnu~is not directly used as a constraint on the stellar
models, it nonetheless enters into the seismic constraint indirectly.
Following \citet{viani_dnu_2019}, for this purpose we used the value of
\Dnu~obtained by fitting a linear relation of the form of
\cref{eq:asymptotic} to the radial p-modes, rather than using values
determined by analysis of the raw power spectrum. For each star we then
derived posterior distributions from these cost functions using the same
Bayesian Monte-Carlo grid-search procedure as in
\citet{ong_differential_2021}, except for the following specific
modifications required to accommodate subgiants in particular.

\hypertarget{frequencies-coupling-matrices-and-surface-corrections}{%
\subsection{Frequencies, Coupling Matrices, and Surface
Corrections}\label{frequencies-coupling-matrices-and-surface-corrections}}

\label{sub:surface}

As in \citet{ong_differential_2021}, we compare the \bg~parametrisation
of the surface term, as our fiducial parametrisation, with the
\(\epsilon\)-matching algorithm of \citet{roxburgh_asteroseismic_2016},
which is nonparametric. In addition to directly comparing the two
methods per se, we also compare two different approaches to mode
coupling. As described in Paper I, all extant parametrisations of the
surface term (including that of \bg) account for mode coupling
indirectly, if they do at all, via weighting the surface-induced
frequency perturbation by the mode inertiae: \begin{equation}
    \delta\nu_{\text{surf},nl} \sim f(\nu_{nl}) / I_{nl}.\label{eq:firstorder}
\end{equation} In Paper I, we showed that expressions of this form
correspond to truncating a perturbative series describing the coupling
between the \(p\)- and \(g\)-mode cavities, with respect to the natural
basis of \(\pi\)- and \(\gamma\)-modes
\citep[in the sense of][]{aizenman_avoided_1977,ong_semianalytic_2020},
to first order in the coupling strengths. In this construction the
surface term is understood to result from the action of some
undetermined structural perturbation operator, which is annihilated on
the \(\gamma\)-mode subspace. The approximations that are implicit in
this sort of truncation are not always valid; Paper I demonstrates that
they hold good for young subgiants with strong coupling, and fail for
very evolved red giant stars. Without any a priori constraints on the
nature of the surface modelling error, however, it is difficult to say
for sure whether or not these approximations hold good on the sample of
subgiants under consideration here.

We thus compare the first-order approach to mode coupling,
\cref{eq:firstorder}, with the use of the full matrix construction
presented in \citet{ong_semianalytic_2020}. Paper I provided the full
details of the numerical framework required to perform this computation,
and derived generalisations of both the \bg~parametrisation and the
\(\epsilon\)-matching technique which account fully for mode coupling.
For this work, we will compare the following surface corrections:

\begin{enumerate}
\def\labelenumi{\arabic{enumi}.}
\tightlist
\item
  The two-term parametric correction of \bg: \begin{equation}
  \delta\nu_{\text{surf},nl} \sim \left.\left(a_{-1}\left(\nu_{nl} \over \nu_0)\right)^{-1} + a_{3}\left(\nu_{nl} \over \nu_0)\right)^{3}\right) \right/ I_{nl}.\label{eq:bg14}
  \end{equation} To ensure that the constraints are only determined by
  the p-mode subsystem, the parameters \(a_{-1}\) and \(a_3\) are fitted
  against only the radial modes, as in \citet{basu_robustness_2018}.
  However, the cost function is evaluated against all modes (including
  mixed modes).
\item
  Using the same parameters, we also evaluate a cost function applying a
  matrix-based generalisation of the \bg~correction. We compute a
  \(\chi^2\) statistic with contributions from the corrected radial and
  quadrupole \(p\)-modes using \cref{eq:bg14}, but for the dipole modes
  we compute corrected frequencies fully accounting for mixed-mode
  coupling (via eqs. 8 and 18 of Paper I).
\item
  We also compute a nonparametric cost function by
  \(\epsilon\)-matching, accounting for mode mixing to first order as
  described in section 3.1 of Paper I.
\item
  Finally, we perform \(\epsilon\)-matching while fully accounting for
  mode coupling, as described in section 3.3 of Paper I.
\end{enumerate}

For this purpose, we computed both the \(\pi\) and \(\gamma\) mode
frequencies and mixed-mode coupling matrices via the mode isolation
construction of \ob, as well as the mixed-mode frequencies themselves,
for p-modes within \(\pm6\Dnu\) of \numax. These computations were done
using the stellar pulsation code \gyre~\citep{townsend_gyre_2013}. We
used only the mixed-mode and \(\pi\)-mode frequencies for surface
corrections which account for mode coupling to first order (1 and 3). In
order to address issues of matrix completeness where the matrix
construction was required (2 and 4), we compute \(\gamma\)-mode
frequencies and matrix elements for \(\gamma\)-modes from a frequency of
\(\numax - 7\Dnu\) all the way up to the \(n_g=1\) \(\gamma\)-mode,
whose frequency is bounded from above by the maximum value of the
Brunt-Väisälä frequency \(N\) as
\(\nu_{n_g=1} < \max_{r < R} (N/2\pi)\). Since this is potentially much
higher than the highest-frequency p-mode in the observational range, we
restricted the effective matrix to omit the highest-frequency \(n_0\)
\(\gamma\)-modes, where \(n_0\) was chosen to minimise the sum of
squared differences between the mixed modes returned from the coupling
eigenvalue equation, and those returned directly from \gyre. This was
only done when the \(n_g=1\) \(\gamma\)-mode was higher in frequency
than the highest \(\pi\)-mode frequency in the incomplete coupling
matrix.

The surface-corrected frequencies were then used to compute a cost
function of the form \begin{equation}
    \chi^2_\text{seis} = {1 \over N_\nu}\sum_n^{N_\nu} \left(\nu_{\text{obs}, n} - \nu_{\text{corr}, n} \over \sqrt{\sigma_{\nu, n}^2 + \sigma_{\nu, \text{eff}}^2}\right)^2,\label{eq:seis}
\end{equation} where \(\sigma_{\nu, \text{eff}}\) quantifies the
systematic error in the modelling procedure owing to grid undersampling
(see the next section).

As the computation of the corrected mode frequencies using the
coupling-matrix eigenvalues is considerably more expensive than the
first-order computation, we restricted this computation to only models
which were both within a spectroscopic Mahalanobis distance of 25
(i.e.~within 5\(\sigma\)) from the nominal spectroscopic constraints,
and also with \(\Dnu\) and \(\numax\) of within 25\% of the
observational values. For each star, we assigned models outside of this
region a likelihood of 0.

\hypertarget{undersampling-of-parameter-space}{%
\subsection{Undersampling of Parameter
Space}\label{undersampling-of-parameter-space}}

Subgiants pose a significant challenge for grid-based modelling as the
evolution of their mode frequencies through avoided crossings is very
rapid compared to the changes undergone by their pure p and g-mode
frequencies, considered separately \citep{deheuvels_constraints_2011}.
These avoided crossings also amplify systematic dependences of the
inferred stellar properties on model parameters like the convective
mixing length. At this time, explicitly increasing the sampling density
of model grids to match the typical frequency measurement errors remains
prohibitively expensive to perform at large scale.

For p-modes, such as the pure p-modes seen in main-sequence stars, or
p-dominated mixed modes in red giants, the effective parameter sampling
density of model grids can be significantly improved by multivariate
interpolation \cite[e.g.][]{rendle_aims_2019}. Numerical stability in
this interpolation problem is achieved by constructing interpolants for
the dimensionless phase quantities \(\epsilon_{l,n}\) of
\cref{eq:asymptotic}, rather than the mode frequencies themselves. By
doing so, the interpolants better describe subtle changes in internal
structure reflected in individual modes. This is as opposed to changes
in mode frequencies that result from homology transformations, which
instead affect all modes by rescaling the overall characteristic
frequency \Dnu, which in turn may be interpolated separately.

Since avoided crossings in subgiants are not well-described by a single
characteristic frequency scale, their frequencies cannot be
preconditioned for interpolation in this manner. Interpolation of
subgiant models therefore requires model grids of already high sampling
density, as shown in \citet{li_ages_2020}. Moreover, the characteristic
frequencies of both the p- and g-mode cavities each evolve rapidly as
the subgiants cross the Hertzsprung gap. Consequently, while very high
temporal and parameter sampling are required to match the strong
structural constraints imposed by typical observational precision on the
mode frequencies, conventional methods of increasing the effective grid
sampling density fail in this regime of rapid evolution.

\citet{li_ages_2020} treat this by systematically inflating the
effective measurement errors associated with the mode frequency
measurements to match the parameter sampling of the grid. This
effectively downweights the seismic cost term in \cref{eq:chi2} relative
to the classical spectroscopic constraints. We adopt a similar measure
in our procedure. For each star, we first identify the best-fitting
model in the grid, using \cref{eq:seis} with \(\sigma_{\nu,\text{eff}}\)
set to zero. We then estimate the effective systematic frequency error
\(\sigma_{\nu,\text{eff}}\) as the root-mean-square difference between
the surface-corrected and observed mode frequencies for this
best-fitting model in the grid, and add this in quadrature to the
nominal frequency measurement errors when evaluating the posterior
distribution functions in all subsequent calculations. Unlike
\citet{li_ages_2020}, we use a single value of this undersampling error
for modes of all degrees \(l\).

\hypertarget{regularisation}{%
\subsection{Regularisation}\label{regularisation}}

Finally, it is a known property of the surface term that the differences
between observed and model p-mode frequencies should decrease in
magnitude at low frequencies, for models with matching interior
structures to the corresponding stars. \citet{basu_robustness_2018} note
that the seismic likelihood functions from different surface-term
treatments may not adequately penalise stellar models in the grid which
do not satisfy this property, necessitating further regularisation. As
in \citet{basu_robustness_2018} and \citet{ong_differential_2021}, we
introduce an additional penalty function \begin{equation}
    \chi^2_\text{low $n$} = {1 \over N}\sum_n^N \left(\nu_{\text{obs}, n} - \nu_{\text{mod}, n} \over \sqrt{\sigma_{\nu,n}^2 + \sigma_{\nu, \text{eff}}^2}\right)^2,
\end{equation} over the lowest \(N\) uncorrected model frequencies. For
this work we choose \(N=4\). Since this contribution to the cost
function only serves the purpose of regularisation, and is not meant to
significantly influence the overall posterior distribution, we
downweight it by a factor of 100.

\hypertarget{subgiant-sample}{%
\section{Subgiant Sample}\label{subgiant-sample}}

\label{sec:sample}

Mode frequencies of subgiants exhibit a broad range of observational
characteristics, as their mixed modes span the transition from being
largely p-dominated and interrupted by a sparse set of g-modes, to being
largely g-dominated and interrupted by a sparse set of p-modes. The
transition between these two regimes is described by the ratio between
the characteristic frequency spacings of the p-mode and g-mode cavities,
\(\Delta\nu / \numax^2 \Delta P\). This quantity exhibits a strong
dependence on \Dnu. As such, for this study we model subgiants across a
large range of \Dnu. At high \Dnu~we have stars barely off the main
sequence, whose subgiant status we diagnose with the presence of
individually identifiable avoided crossings. At lower frequencies, we
have stars part way up the red giant branch, where we adopt a minimum
\Dnu~of \(15\ \mu\)Hz for inclusion in our sample. Additionally, we
include in our sample stars which were observed with different space
missions. In total, our subgiant sample consists of 36 stars observed
with \textit{Kepler}, 6 stars observed with \textit{TESS}, and 5
subgiants observed with \textit{K2}. We describe each of these
subsamples in more detail below.

\hypertarget{kepler-subgiants}{%
\subsection{\texorpdfstring{\emph{Kepler}
Subgiants}{Kepler Subgiants}}\label{kepler-subgiants}}

Photometry from the nominal \emph{Kepler} mission, spanning its full
duration of just over 4 years, represents the best possible scenario for
photometric measurements of oscillatory variability in solar-like
pulsators. The long duration and high completeness of these time series
ensure high resolution in their power spectra with relatively clean
spectral line-spread functions, while the comparatively high sampling
frequency of short-cadence observations (which is much higher than
\numax) reduces signal apodisation when measuring the oscillations of
subgiants and main-sequence stars.

For this work, we consider a sample of subgiants observed at short
cadence which has previously been subject to grid-based modelling with
detailed seismic constraints \citep{li_ages_2020, li_frequencies_2020}.
For these targets, we use the global seismic properties \Dnu~and
\numax~as derived by \citet{serenelli_apokasc_2017}, and the
consolidated spectroscopic observables (\(\teff\) and \(\feh\)) from
Table 1 of \citet{li_ages_2020}. For detailed modelling, we also use the
mode frequencies measured in \citet{li_frequencies_2020}.

\hypertarget{tess-subgiants}{%
\subsection{\texorpdfstring{\emph{TESS}
Subgiants}{TESS Subgiants}}\label{tess-subgiants}}

\begin{table*}[htbp]
\centering
  \caption{Consolidated spectroscopic and seismic constraints for TESS subgiants. \label{tab:tess}}
  \tabularnewline
  \begin{tabular}{cccccccccccc}
  \toprule
  Object & \teff/K & \(\sigma_{\teff}\)/K & \(\feh\) (dex) & \(\sigma_{\feh}\) (dex) & \(L/L_\Sun\) & \(\sigma_{L}/L_\Sun\) & \(\Dnu/\mu\)Hz & \(\sigma_{\Dnu}/\mu\)Hz & \(\numax/\mu\)Hz & \(\sigma_{\numax}/\mu\)Hz & Reference \\
  \midrule
  HD 38529 & 5578 & 52 & \(+0.34\) & \(0.06\) & 6.233 & 0.149 & 36.68 & 0.82 & 624 & 20 & 1 \\
  \(\beta\) Hyi & 5874 & 74 & \(-0.10\) & \(0.09\) & 3.494 & 0.087 & 57.85 & 0.15 & -- & -- & 2 \\
  \(\nu\) Ind & 5320 & 64 & \(-1.18\)\footnote{Corrected for $\alpha$-enrichment by prescription of \citet{salaris_alpha_1993}} & \(0.11\) & 6.000 & 0.350 & 25.10 & 0.10 & -- & -- & 3 \\
  \(\delta\) Eri & 4954 & 30 & \(+0.06\) & \(0.05\) & -- & -- & 40.58 & 0.14 & 669 & 7 & 4 \\
  TOI 197 & 5080 & 90 & \(-0.08\) & \(0.08\) & 5.150 & 0.170 & 28.94 & 0.15 & 430 & 18 & 5 \\
  \(\eta\) Cep & 4970 & 100 & \(-0.09\){\textsuperscript{a}} & \(0.11\) & 8.130 & 0.290 & 17.23 & 0.05 & 228 & 2 & 6 \\
  \bottomrule
  \end{tabular}

  \begin{flushleft}
  \footnotesize
  References: 1. \citet{ball_robust_2020}; 2. White et al.~in prep.; 3. \citet{chaplin_age_2020}; 4. Bellinger et al.~in prep.; 5. \citet{huber_saturn_2019}; 6. Joyce et al., submitted
  \end{flushleft}
\end{table*}

We also consider a sample of subgiants observed with TESS during its
nominal mission. This represents the opposite extreme of very short time
series, where typically only a handful of modes can be detected (if at
all). In the worst case, some of these subgiants have continuous
photometry spanning only a single sector (27 days), which is not
appreciably longer than the average mode lifetimes. Equivalently, in the
frequency domain, the spectral resolution of the power spectrum --- and
so the correlation frequency scale of the underlying correlated
realisation noise --- is comparable to the excited mode linewidths.
Consequently, the power spectra and fitted frequencies of these targets
are far more sensitive to realisation noise than the \emph{Kepler}
sample. To make matters worse, TESS pixels are significantly larger than
\emph{Kepler} ones, and therefore potentially suffer more contamination
from nearby objects. Generically speaking, the detection and extraction
of seismic characteristics from these short and relatively contaminated
power spectra --- let alone individual mode frequencies --- is a
challenging task.

Since photometric and time-series data analysis in this very difficult
regime is not the focus of this work, we restrict our attention to only
a relatively small sample of TESS subgiant oscillators for which mode
frequencies have previously been identified by and circulated within the
asteroseismology community. These are TOI-197 \citep{huber_saturn_2019},
\(\nu\) Ind \citep{chaplin_age_2020}, \(\beta\) Hyi (White et al.~in
prep.), \(\delta\) Eri (Bellinger et al.~in prep.), HD 38529
\citep{ball_robust_2020}, and \(\eta\) Cep (Joyce et al.~submitted). We
show their consolidated spectroscopic and global seismic properties in
\cref{tab:tess}.

\hypertarget{k2-subgiants}{%
\subsection{K2 Subgiants}\label{k2-subgiants}}

Data from the K2 mission lie in between these two extremes. On one hand,
the time series available from K2 span 80 days in each campaign (as
opposed to 27 days in a TESS sector), with 60-second sampling at short
cadence (as opposed to TESS's 2 minutes). On the other hand, the
pointing stability (and so photometric noise) of these time series is
much degraded compared to the nominal Kepler mission, owing to the
drift-scanning strategy across the ecliptic plane adopted by K2,
necessitated by loss of attitude control. For our sample, we use a set
of K2 short-cadence stars observed across multiple campaigns for which a
significant oscillation power excess could be identified by visual
inspection of their photometric power spectra. The oscillation
frequencies and spectroscopic properties of these stars have not been
previously published; we describe them in detail below.

\hypertarget{spectroscopic-constraints}{%
\subsubsection{Spectroscopic
Constraints}\label{spectroscopic-constraints}}

Single spectra for each of these targets were obtained between UT 2017
June 09 and June 12, using the Tillinghast Reflector Echelle
Spectrograph \citep[TRES, see][]{gaborthesis, TRES} on the 1.5m
telescope at the Fred L. Whipple Observatory (FLWO) on Mt. Hopkins in
Arizona. TRES is an optical echelle spectrograph with a wavelength range
385-910 nm and a resolving power of \(R \sim 44,000\). The TRES spectra
were extracted using procedures described in \citet{buchhave_hat_2010},
and then used to derive stellar effective temperatures (\teff),
metallicities ({[}M/H{]}), and surface gravities (\logg) using the
Stellar Parameter Classification tool \citep[SPC,
see][]{buchhave_abundance_2012} without any stellar isochrone models as
a prior. SPC cross-correlates an observed spectrum against a grid of
synthetic spectra based on Kurucz atmospheric models
\citep{kurucz_model_1992}.

For each star, an initial set of spectroscopic parameters was derived by
permitting each parameter to vary freely in the fit. As done in
\citet{lund_properties_2016}, the resulting value of \teff~was then used
to calculate a new value of the surface gravity consistent with
\numax~obtained from their photometric power spectra (described in the
following section), via the solar-calibrated \numax~scaling relation
\citep{kjeldsen_amplitudes_1995} \begin{equation}
    \numax \sim \left(M \over M_\Sun\right) \left(R \over R_\Sun\right)^{-2} \left(\teff \over \teff_\Sun\right)^{-{1\over2}} \numax_\Sun.\label{eq:numax}
\end{equation} This seismic value of \logg~\citep[calibrated against the
solar value of \(g = 27402\ \mathrm{cm\ s^{-1}}\),
from][]{campante_limits_2014} was then used to perform another SPC
analysis, run with \logg~as a fixed parameter, yielding a final set of
spectroscopic values.

\hypertarget{power-spectra-and-seismic-constraints}{%
\subsubsection{Power Spectra and Seismic
Constraints}\label{power-spectra-and-seismic-constraints}}

We use power spectra derived from photometric time series, which were in
turn processed using the K2P\textsuperscript{2} pipeline
\citep{lund_k2p2_2015} with custom apertures in order to ameliorate CCD
bleeding, and corrected using the KASOC filter
\citep{handberg_automated_2014}. Weighted Lomb-Scargle power spectra
were computed for each campaign separately out of the
K2P\textsuperscript{2} time series, with each point in the time series
weighted by the root-mean-square photometric variability within a
symmetric 2-day running filter. A final power spectrum was then made
from a weighted average of the power spectra from individual campaigns.
For a more detailed description, see Lund et al., in prep.

From these combined power spectra, the global seismic parameters
\Dnu~and \numax~were found using the CV method of \citet{bell_cv_2019}.
The value of \Dnu~found here was only used as an initial estimate to
seed the mode identification for peakbagging; for modelling purposes we
refit \Dnu~as described above. Mode frequencies were then fitted against
these power spectra using a combination of techniques. Radial and
quadrupole modes were identified and fitted using PBJam \citep{pbjam};
dipole modes were identified by hand by visual inspection of the
replicated échelle diagrams \citep{bedding_replicated_2012} and fitted
using \diamonds~\citep{diamonds}. For modelling we retained only modes
with a peak signal-to-noise ratio larger than 25.

We show the consolidated spectroscopic and seismic properties of these
K2 stars in \cref{tab:k2spec}, and their mode frequencies in
\cref{tab:k2seis} in the Appendix. Of the 11 stars in the original
sample, 4 stars were main-sequence stars (with \(\Dnu\) between
\(65\ \mu\)Hz and \(133\ \mu\)Hz and no discernible mode mixing), and 2
stars were red giants (\(\Dnu < 15\ \mu\)Hz). We excluded these from
consideration for our grid-based modelling.

\begin{longtable*}[]{@{}
  >{\raggedright\arraybackslash}p{(\columnwidth - 16\tabcolsep) * \real{0.20}}
  >{\centering\arraybackslash}p{(\columnwidth - 16\tabcolsep) * \real{0.10}}
  >{\centering\arraybackslash}p{(\columnwidth - 16\tabcolsep) * \real{0.13}}
  >{\centering\arraybackslash}p{(\columnwidth - 16\tabcolsep) * \real{0.10}}
  >{\centering\arraybackslash}p{(\columnwidth - 16\tabcolsep) * \real{0.10}}
  >{\centering\arraybackslash}p{(\columnwidth - 16\tabcolsep) * \real{0.10}}
  >{\centering\arraybackslash}p{(\columnwidth - 16\tabcolsep) * \real{0.10}}
  >{\centering\arraybackslash}p{(\columnwidth - 16\tabcolsep) * \real{0.10}}
  >{\raggedright\arraybackslash}p{(\columnwidth - 16\tabcolsep) * \real{0.08}}@{}}
\caption{Consolidated spectroscopic and seismic constraints for our K2
sample. We adopted a uniform value of \(\sigma_\teff=50\ \mathrm{K}\)
and \(\sigma_\text{[M/H]} = 0.08\) dex. Luminosities are from
\cite{gaia_dr2_2018}. \label{tab:k2spec}}\tabularnewline
\toprule
\begin{minipage}[b]{\linewidth}\raggedright
EPIC
\end{minipage} & \begin{minipage}[b]{\linewidth}\centering
\teff~/K
\end{minipage} & \begin{minipage}[b]{\linewidth}\centering
\(\text{[M/H]}\) (dex)
\end{minipage} & \begin{minipage}[b]{\linewidth}\centering
\(\Dnu\) /\(\mu\)Hz
\end{minipage} & \begin{minipage}[b]{\linewidth}\centering
\(\sigma_{\Dnu}\) /\(\mu\)Hz
\end{minipage} & \begin{minipage}[b]{\linewidth}\centering
\(\numax\) /\(\mu\)Hz
\end{minipage} & \begin{minipage}[b]{\linewidth}\centering
\(\sigma_\numax\) /\(\mu\)Hz
\end{minipage} & \begin{minipage}[b]{\linewidth}\centering
\(L\) /\(L_\odot\)
\end{minipage} & \begin{minipage}[b]{\linewidth}\raggedright
\(\sigma_L\) /\(L_\odot\)
\end{minipage} \\
\midrule
\endfirsthead
\toprule
\begin{minipage}[b]{\linewidth}\raggedright
EPIC
\end{minipage} & \begin{minipage}[b]{\linewidth}\centering
\teff~/K
\end{minipage} & \begin{minipage}[b]{\linewidth}\centering
\(\text{[M/H]}\) (dex)
\end{minipage} & \begin{minipage}[b]{\linewidth}\centering
\(\Dnu\) /\(\mu\)Hz
\end{minipage} & \begin{minipage}[b]{\linewidth}\centering
\(\sigma_{\Dnu}\) /\(\mu\)Hz
\end{minipage} & \begin{minipage}[b]{\linewidth}\centering
\(\numax\) /\(\mu\)Hz
\end{minipage} & \begin{minipage}[b]{\linewidth}\centering
\(\sigma_\numax\) /\(\mu\)Hz
\end{minipage} & \begin{minipage}[b]{\linewidth}\centering
\(L\) /\(L_\odot\)
\end{minipage} & \begin{minipage}[b]{\linewidth}\raggedright
\(\sigma_L\) /\(L_\odot\)
\end{minipage} \\
\midrule
\endhead
212478598 & \(5058\) & \(-0.356\) & 35.09 & 0.05 & 536.7 & 4.8 & --- &
--- \\
212485100 & \(6169\) & \(-0.041\) & 87.34 & 0.12 & 1834.3 & 21.4 & --- &
--- \\
212487676 & \(6036\) & \(-0.314\) & 75.89 & 0.21 & 1448.2 & 14.5 & 2.651
& 0.023 \\
212516207 & \(6253\) & \(+0.130\) & 68.02 & 0.26 & 1373.6 & 20.0 & 4.031
& 0.042 \\
212586030 & \(4895\) & \(+0.230\) & 21.70 & 0.03 & 311.5 & 10.0 & 6.079
& 0.273 \\
212683142 & \(5898\) & \(-0.052\) & 45.76 & 0.05 & 806.1 & 8.3 & 5.670 &
0.052 \\
212708252 & \(5616\) & \(-0.095\) & 132.95 & 0.26 & 2896.7 & 21.7 &
0.887 & 0.003 \\
246154489 & \(5018\) & \(-0.355\) & 14.28 & 0.03 & 187.5 & 4.4 & 12.38 &
0.12 \\
246184564 & \(4932\) & \(-0.131\) & 11.82 & 0.02 & 155.4 & 5.4 & 16.94 &
0.25 \\
246305274 & \(6178\) & \(-0.431\) & 46.60 & 0.24 & 795.2 & 12.3 & --- &
--- \\
246305350 & \(5991\) & \(+0.011\) & 48.00 & 0.18 & 831.7 & 13.6 & 5.767
& 0.051 \\
\bottomrule
\end{longtable*}

\hypertarget{results}{%
\section{Results}\label{results}}

\label{sec:results}

\begin{figure}
\centering
\includegraphics{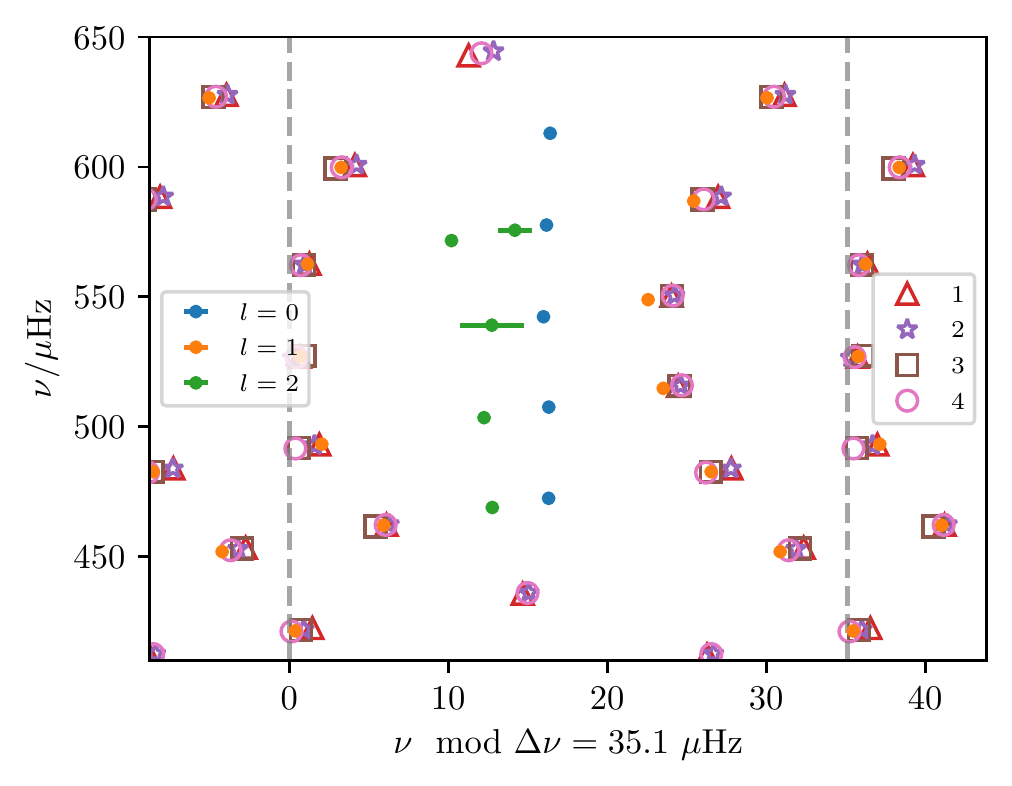}
\caption{Replicated echelle diagram of EPIC 212478598, showing observed
modes as points coloured by angular degree, and model \(l=1\)
frequencies corrected with various prescriptions for the surface term
with the open symbols. With respect to the descriptions in
\autoref{sub:surface}: we show correction 1 with the upright triangles,
2 with the stars, 3 with the squares, and 4 with the open circles.
\label{fig:echelle}}
\end{figure}

For stellar models which provide good matches to the internal structure
of the observed stars, we find that the differences between the
corrected frequencies implied by each of the surface-term treatments we
have considered here can be quite subtle. We show a representative
example in \cref{fig:echelle}, which is the best-fitting model for
EPIC212478598 with respect to the first-order \bg~correction. In the
neighbourhood of \numax, the different corrections yield (at least
visually) very similar results on the dipole mixed modes. Variations in
the posterior distributions do not originate from how they respond to
surface-localised structural differences, as by construction such
frequency errors should lie mainly in the nullspace of all of these cost
functions if the surface term is small. Instead, they result from
``false positives'': differences in how these prescriptions might fail
to penalise interior structural differences between the star and the
model. It is difficult to say a priori, from strictly theoretical
considerations, how the inferred stellar properties might respond to
these differences, all else being equal. This is why explicit
comparisons of the inferred properties are required.

\begin{table*}[htbp]
\centering
    \caption{$t$-test $p$-values for differences between modelling methodologies, under the null hypothesis that the normalised differences $z$, \cref{eq:z}, are distributed with a population mean of $0$. If $p < \alpha = 0.0025$, then estimates of a given property returned by 2 different methods are on average not consistent with each other at the $3\sigma$ level; these are indicated with bold face.\label{tab:pvalues}}
    \tabularnewline
    \begin{tabular}{ c | c c | c c c | c c | c }
    \toprule
    Quantity& \multicolumn{2}{c}{First-order vs. Full} & \multicolumn{3}{c}{BG14 vs. $\epsilon$-matching} &  \multicolumn{2}{c}{$L$ vs. no $L$} & This work vs. \cite{li_ages_2020} \\
    & BG14 {\scriptsize(1 vs. 2)} & $\epsilon$-match \scriptsize(3 vs. 4) & 1\textsuperscript{st}-order \scriptsize(1 vs. 3) & Full \scriptsize(2 vs. 4) & Full with $L$ & BG14 \scriptsize(2) & $\epsilon$-match \scriptsize(4) & BG14 {\scriptsize(1)}, without $L$\\
    \midrule
    Mass & $\bf 7.588 \times 10^{-5}$ & \bf 0.0009 & 0.3473 & \bf 0.0009 & 0.4787 & 0.9921 & 0.0851 & \bf 0.0016 \\
    Radius & 0.0139 & 0.0267 & 0.0150 & 0.0220 & 0.3061 & 0.6420 & 0.0044 & 0.0445 \\
    Age & 0.8918 & 0.7374 & \bf 0.0003 & 0.0037 & 0.9041 & 0.9008 & 0.0371 & 0.0094 \\
    $Y_0$ & \bf 0.0024 & 0.0258 & 0.1117 & 0.3904 & 0.0754 & 0.1526 & 0.9241 & - \\
    \bottomrule
    \end{tabular}
\end{table*}

Following \citet{ong_differential_2021}, we compare the different
modelling methods by way of their normalised differences between them.
For a given property (e.g.~the stellar mass \(M\)), we define a
normalised score, e.g.~\(z_M\), as the differences between the posterior
mean values \(\mu_M\) as reported by two different estimation methods
\(A\) and \(B\), normalised by the combination in quadrature of their
reported uncertainties, for which we use their posterior variances
\(\sigma^2_M\): \begin{equation}
    z_{M, A\text{ vs. } M, B} = {\mu_{M,A} - \mu_{M,B} \over \sqrt{\sigma^2_{M,A} + \sigma^2_{M,B}}}.\label{eq:z}
\end{equation} Under the null hypothesis that estimation methods \(A\)
and \(B\) are consistent with each other, the distribution of these
\(z\)-scores should be distributed with a population mean of 0. Our
alternative hypothesis, that \(A\) and \(B\) are not consistent with
each other, does not constrain the sign of the population mean of \(z\).
Consequently, we perform one-sample 2-tailed \(t\)-tests of this null
hypothesis, following \citet{silvaaguirre_standing_2017} and
\citet{ong_differential_2021}. As in \citet{ong_differential_2021} we
adopt a \(3\sigma\) level of significance (i.e.~we reject the null
hypothesis for \(p \lesssim 0.0025\)). We perform comparisons of this
kind on some fundamental stellar properties which are of astrophysical
interest: the stellar masses, radii, ages, and initial helium
abundances. We show the \(p\)-values associated with various comparisons
that we performed in \cref{tab:pvalues}. In the next few sections, we
explain each set of comparisons in detail.

\hypertarget{full-vs.-first-order-mode-coupling}{%
\subsection{Full vs.~First-order Mode
Coupling}\label{full-vs.-first-order-mode-coupling}}

The first two columns of \cref{tab:pvalues} describe comparisons between
theoretical constructions of the surface term that are identical save
for how mode coupling is handled. In the first column, we compare the
first-order and full matrix generalisation of the \bg~surface term
(corrections 1 and 2 of \autoref{sub:surface}), and in the second we
compare the first-order and full matrix generalisation of the
\(\epsilon\)-matching algorithm (corrections 3 and 4). These correspond
to the pairwise comparisons, and cumulative distribution functions of
\(z\)-scores, shown in \cref{fig:comp1}.

In each of these cases we find that the reported values of the stellar
mass differ significantly within each pair. We also see a similar
(albeit somewhat marginally significant) difference in the inferred
initial helium abundances for the \bg~correction, but not for
\(\epsilon\)-matching. We attribute this discrepancy to
\(\epsilon\)-matching returning somewhat larger posterior uncertainties
in general, since it places considerably weaker constraints on the
functional form of the surface term in consequence of its nonparametric
nature. These differences indicate that explicitly accounting for mode
coupling to beyond first order when computing corrections for the
surface term may be necessary for estimating masses and helium
abundances of evolved stars with mixed modes, even given the present
limitations of grid-based modelling.

We do not obtain any other statistically significant differences of this
kind. This does not necessarily imply that first-order and full
treatments of mode coupling return equivalent results; the only
conclusion we may draw is that even if any such differences exist, the
limitations of both the observational precision and intrinsic systematic
errors in our grid-based modelling are more severe than would permit
them to be demonstrated at the \(3\sigma\) level by the techniques we
have used.

\hypertarget{parametric-vs.-nonparametric-treatment}{%
\subsection{Parametric vs.~Nonparametric
Treatment}\label{parametric-vs.-nonparametric-treatment}}

In the third and fourth columns of \cref{tab:pvalues} we show the
results of comparing different formulations of the surface term, keeping
the treatment of mode coupling the same. In the third column we compare
the surface correction of \bg~against the \(\epsilon\)-matching
algorithm, in both cases only accounting for mode mixing to first order;
in the fourth we compare these methods as computed with respect to the
coupling-matrix eigenvalue equation.

We see significant systematic effects in the inferred masses and ages,
which we show in panels a, b, e, and f of \cref{fig:comp2}. These
results are slightly harder to interpret. While the two formulations
appear not to disagree on the stellar masses at first order in mode
coupling, they nonetheless select models in the grid with different
stellar ages. Since the estimated stellar radii (constrained by \Dnu)
and initial helium abundances (constrained by mixed-mode avoided
crossings) returned by the two methods do not appear to differ
significantly, and both methods are equally constrained by the
spectroscopic metallicity, we may attribute these age variations to
model degeneracies between the age and the internal mixing physics
\citep[e.g.][]{lebreton_alacarte_2014}. In particular, at a fixed mass
and radius, variations in the mixing length parameter \amlt, and the
step overshoot parameter \fov, will result in differing positions of
both the convective-envelope boundary and ionisation-zone acoustic
glitches, and (for intermediate-mass stars with convective
hydrogen-burning cores on the main sequence) the fossil acoustic
signature of the former convective core boundary during the subgiant
phase. As noted in \citet{ong_differential_2021}, these structural
mismatches, localised to the interior, are penalised differently by
various constructions of the surface term. Both of these parameters,
which describe the mixing physics of stellar models, are also known to
affect their main-sequence lifetimes; the main-sequence lifetime of a
stellar model in turn strongly determines its age during the subgiant
and red giant phases of evolution. It is conventionally assumed that
these degeneracies may be broken by the introduction of the luminosity
as an additional spectroscopic constraint
\citep{lebreton_alacarte_2014}; we discuss this in more detail in the
next two subsections.

Once a full accounting for mixed-mode coupling is introduced, the two
methods do disagree on the inferred masses, while the discrepancies in
the inferred ages are no longer significant at the \(3\sigma\) level.
Generally speaking, we find that the use of a complete prescription for
mode coupling results in slightly wider posterior distributions (and
hence diminished effect sizes when normalised by the reported
uncertainties), which may explain the loss of significance in the
differences in the ages. However, the differences in the reported masses
still suggests that the two surface-term treatments respond differently
to extended descriptions of mode coupling. We have no good a priori
explanation for this phenomenon.

\hypertarget{benchmark-inclusion-of-other-constraints}{%
\subsection{Benchmark: Inclusion of Other
Constraints}\label{benchmark-inclusion-of-other-constraints}}

We now seek to compare the size of the systematic variations originating
from different surface term treatments with those arising from other
methodological decisions in the modelling procedure. For example, we
might introduce an additional contribution to the cost function,
\begin{equation}
    \chi^2_{L} = \left(L_\text{obs} - L_\text{mod} \over \sigma_{L}\right)^2,
\end{equation} where \(L\) is the stellar luminosity. For this exercise
we restricted ourselves to the subset of our targets for which
luminosities were available, either from their associated boutique
modelling efforts (for the TESS targets) or from Gaia DR2
\citep{gaia_dr2_2018} for the Kepler and K2 targets. This excluded 2 K2
subgiants, \(\delta\) Eri, and 3 Kepler subgiants from our sample. We
show the results of this test in the sixth and seventh columns of
\cref{tab:pvalues}, restricting ourselves to considering only cases
where mode coupling is fully accounted for. Contrary to expectations, we
see that the introduction of the luminosity constraint does not
significantly modify the estimates of the stellar properties under our
consideration when using the \bg~surface correction; while the effects
are relatively larger for the \(\epsilon\)-matching algorithm (yielding
smaller \(p\)-values), they are nonetheless also not significant at the
\(3\sigma\) level. We examine this more closely in the next subsection.

The discrepancy between how these two surface-term treatments respond to
the introduction of the luminosity constraint can be better examined by
comparing them directly to each other, fully accounting for mode
coupling, which we show in the fifth column of \cref{tab:pvalues}. We
find that the presence of the luminosity constraint brings the two
surface-term treatments into closer agreement with each other, resolving
the statistical tension between them from the previous section.
Specifically, we show the stellar masses and the distribution of their
normalised differences in \cref{fig:comp2}c and g: we see that the
introduction of the luminosity constraint brings the inferred values
from both methods closer to equality, on average. Since the generalised
\bg~masses (i.e.~with full mode coupling) are also less changed upon the
introduction of the luminosity constraint, this suggests that the
differences in behaviour between the two surface-term treatments
discussed in the previous section may ultimately be a result of
nonparametric methods being potentially too flexible (yielding false
positives) in this phase of evolution. Conversely, this might also imply
that the generalised \bg~correction may provide more robust parameter
estimates for subgiants than \(\epsilon\)-matching, in the absence of
luminosity measurements.

\hypertarget{survey-and-single-target-systematics}{%
\subsection{Survey and Single-target
Systematics}\label{survey-and-single-target-systematics}}

Given that we have included subgiants from surveys of varying data
quality, one might question whether or not those systematic differences
which we have found have been driven by outliers, rather than being
statistically representative of subgiants in general. For our most
significantly discrepant set of pairwise differences --- stellar masses
under different treatments of mode coupling (\cref{fig:comp1}a) --- we
see that there are a few potential outliers in our sample. Subdividing
the \bg~masses by originating survey (\cref{fig:comp2}d), we find that
the most extreme deviations from equality are indeed from \textit{K2}
and \textit{TESS}, which (per our previous discussion) are of somewhat
degraded quality compared to nominal Kepler photometry. However, the
distributions of normalised differences (\cref{fig:comp2}h) are not
substantially modified when we restrict our attention to only
\textit{Kepler} subgiants; repeating the two-tailed \(t\)-test with only
the \textit{Kepler} subsample yields \(p = 5.612 \times 10^{-5}\) for
the \bg~correction and \(p = 0.0014\) for \(\epsilon\)-matching, which
are both significant at the \(3\sigma\) level. In general we find that
our findings in the preceding subsections are not substantively changed
when limited in scope to the \textit{Kepler} subsample.

\begin{figure*}[htbp]
    \centering
    \includegraphics[width=\textwidth]{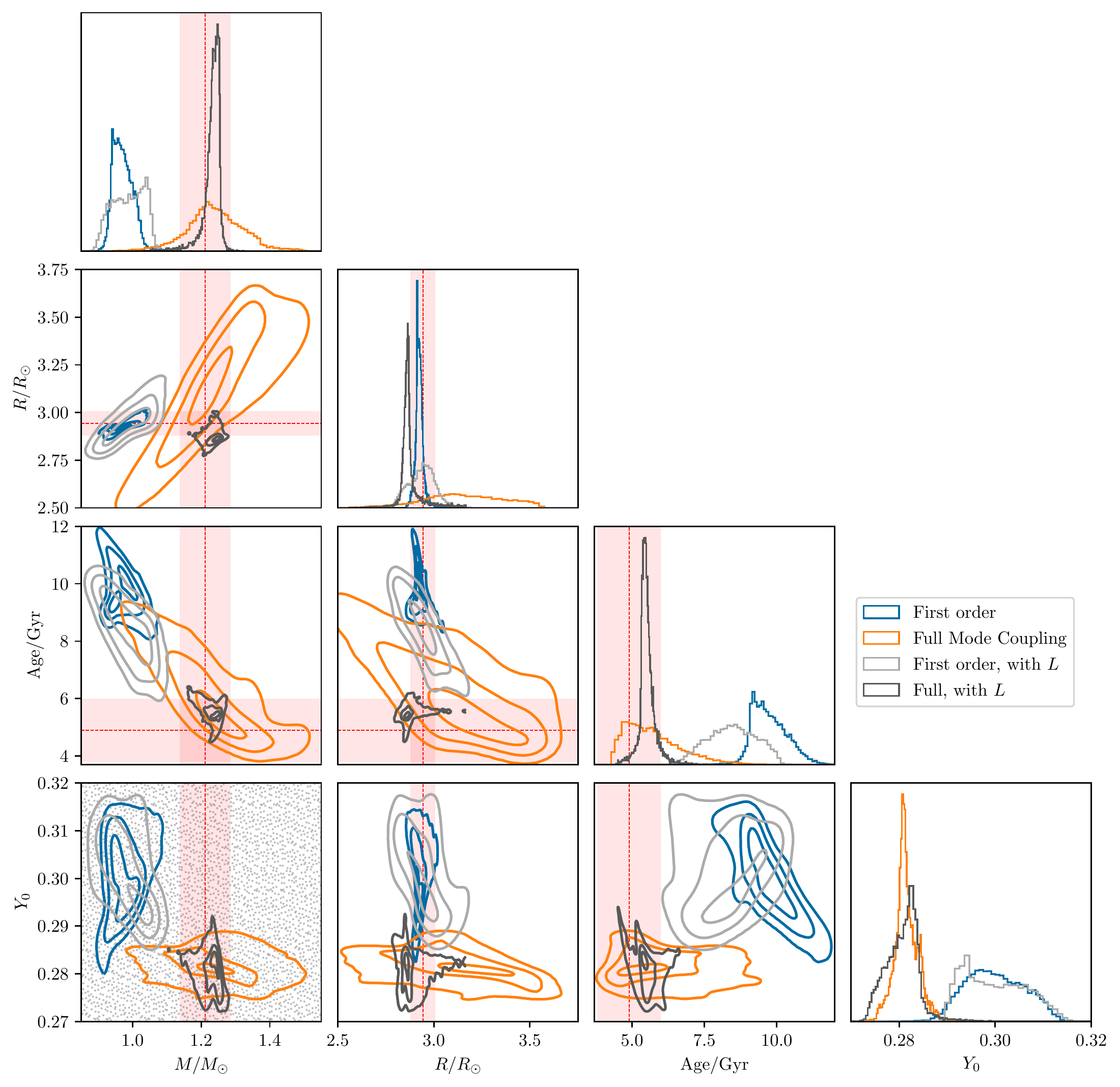}
    \caption{Corner plot for TOI-197, showing the $1\sigma$, $2\sigma$, and $3\sigma$ contours of the joint posterior distributions for different combinations of mode coupling treatment and spectroscopic constraints; the marginal probability density functions are shown on the diagonals. We show with red dashed lines and shaded intervals the literature values and uncertainties for the mass, radius, and age \citep[as obtained from boutique modelling in][]{huber_saturn_2019}. In the bottom-left panel, we show with the grey points the Sobol-sampled values of $M$ and $Y_0$ underlying the grid.}
    \label{fig:toi197}
\end{figure*}

We now examine in detail the most discrepant data point in
\cref{fig:comp2}d, which turns out to be the TESS subgiant TOI-197.
Unusually within our sample, but like most other TESS subgiants, only a
very small number of modes (8 in total with 2 more uncertain) were
identified for this star. Given this paucity of data, g-mode
misidentification \citep[as described in][]{ong_semianalytic_2020} may
yield significant systematic error. Moreover, TOI-197 is in such a state
of evolution that the spacing of dipole g-modes near \numax~is
comparable to that of p-modes, rather than being significantly larger or
smaller. In this regime, a complicated, highly nonasymptotic mixed-mode
pattern is obtained rather than individually identifiable avoided
crossings. As such, extremely fine temporal and parameter sampling
(i.e.~``boutique'' modelling) should be necessary a priori. Fortunately,
\citet{huber_saturn_2019} have already performed just such a boutique
seismic characterisation of TOI-197. In order to account for other
sources of systematic methodological variability (as described in the
next subsection), they employed multiple combinations of parameter
search techniques (both optimisation-based and grid-based), stellar
evolution codes, inputs to stellar evolution (e.g.~chemical mixture),
and input physics. We may therefore assume that their parameter
estimates are generally robust.

We show in \cref{fig:toi197} a comparison of the posterior distributions
obtained using various surface-term treatments (showing the contours of
the first few standard quantiles with the solid coloured curves), in
comparison with the parameter estimates and reported uncertainties of
\citet{huber_saturn_2019}, shown with red lines and shaded intervals. We
compare the posterior distributions for the \bg~correction under
different treatments of mode coupling, with and without the inclusion of
the luminosity constraint. We see that when mode coupling is only
included to first order, the resulting parameter estimates are in
tension with \cite{huber_saturn_2019}, while the inclusion of the
luminosity constraint does not tighten the posterior distribution. On
the contrary, the posterior distribution is broadened, and developes a
lobe in the direction of the nominal values from
\cite{huber_saturn_2019}; this is most visible in the mass-age and
radius-age joint distributions. In the bottom-left panel of
\cref{fig:toi197}, we also show with the grey points the values of the
mass-helium Sobol-sequence samples underlying the grid; this shows that
the peak of the joint distribution is much further away from the nominal
values of \cite{huber_saturn_2019} than can be solely explained by
insufficiently dense sampling of the grid parameters. On the other hand,
when mode coupling is fully accounted for, the posterior distribution
is, while broad, nonetheless consistent with the boutique-modelling
estimate from \cite{huber_saturn_2019}. The introduction of the
luminosity constraint then significantly tightens the posterior
distribution.

Taken together, these are suggestive of the following interpretation:
when fully accounting for mode coupling in the \bg~correction, the
best-fitting region of parameter space sampled by the grid is in good
agreement with the luminosity constraint. The luminosity likelihood
function takes a maximum value near those of the other constraints,
resulting in a considerably tighter posterior distribution when they are
combined. Conversely, the best-fitting region in our grid under the
first-order \bg~correction is in tension with the luminosity constraint.
Moreover, the best-fitting models there lie in a local minimum in the
cost-function landscape, which is sufficiently deep as to interfere with
the luminosity constraint.

Since qualitatively more correct estimates are returned from the full
treatment of mode coupling, we find it unlikely that the difference
between it and the first-order treatment are solely a result of poor
data quality or grid artifacts. It is true that grid searches using an
undersampled grid in combination with a paucity of mode frequencies are
susceptible to spurious local minima, particularly when the
characteristic avoided-crossing pattern
\cite[e.g.][]{bedding_replicated_2012} or reliable g-mode identification
\citep[e.g.][]{ong_semianalytic_2020} are not available. However, this
affects both treatments of mode coupling, and therefore cannot
adequately explain the differences between them. Rather, as we showed in
Paper I, first-order approximations to mode coupling become increasingly
unreliable (for a fixed relative size of the surface term) as stars
evolve to lower values of \Dnu. Consequently, an incomplete treatment of
mode coupling will cause any local minimum in the cost function to lie
in a different region of parameter space, compared to when a correct
accounting of mode coupling is used; the distance between the two will
be larger for more evolved subgiants.

As discussed earlier, luminosity measurements are in principle capable
of resolving the degeneracy between stellar mass and initial helium
abundance. However, we see from \cref{fig:toi197} that in the case of
TOI-197, the treatment of mixed-mode coupling has a more significant
effect on the estimated initial helium abundance than does the presence
or absence of a luminosity constraint. Our above discussion in turn
implies that accurate estimation of the initial helium abundance from
subgiants which are at least as evolved as TOI-197
(i.e.~\(\Dnu \lesssim 30\ \mu\)Hz) will require full mode-coupling
computations to be performed when correcting for the surface term.

\hypertarget{benchmark-other-methodological-variations}{%
\subsection{Benchmark: Other Methodological
Variations}\label{benchmark-other-methodological-variations}}

Finally, we examine methodological variability induced by differences in
how the stellar models underlying the grid itself were constructed. For
this purpose we use the stellar properties as estimated by
\citet{li_ages_2020}. Unlike the one used here, the model grid used in
that work uses a fixed value of the mixing-length parameter, a different
chemical mixture \citep{agss09}, a different sampling scheme
(even-tempered uniform sampling rather than Sobol sequences), a fixed
metallicity-helium relation, and does not account for mass-dependent
core overshoot. Consequently, we would expect offsets in the inferred
ages and masses returned by both of these modelling pipelines, even
adopting identical choices in the surface-term correction. To ensure at
least some consistency, we adopt the same set of constraints: the
spectroscopic observables \teff~and \feh, and the individual mode
frequencies under the \bg~surface correction, taken to only first order
in the mixed-mode coupling.

\begin{figure}[htbp]
    \centering
    \annotate{\includegraphics[trim=.25cm .25cm .25cm .15cm,clip]{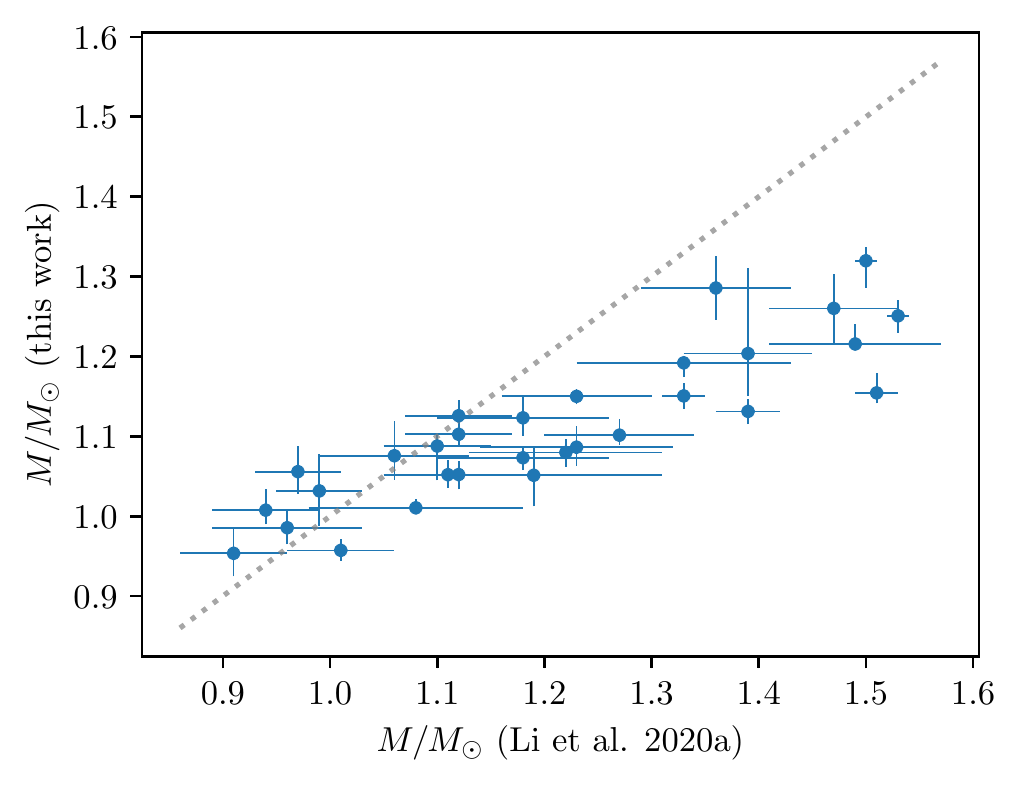}}{\node at (.2, .9){\textbf{(a)}};}
    \annotate{\includegraphics[trim=.25cm .25cm .25cm .15cm,clip]{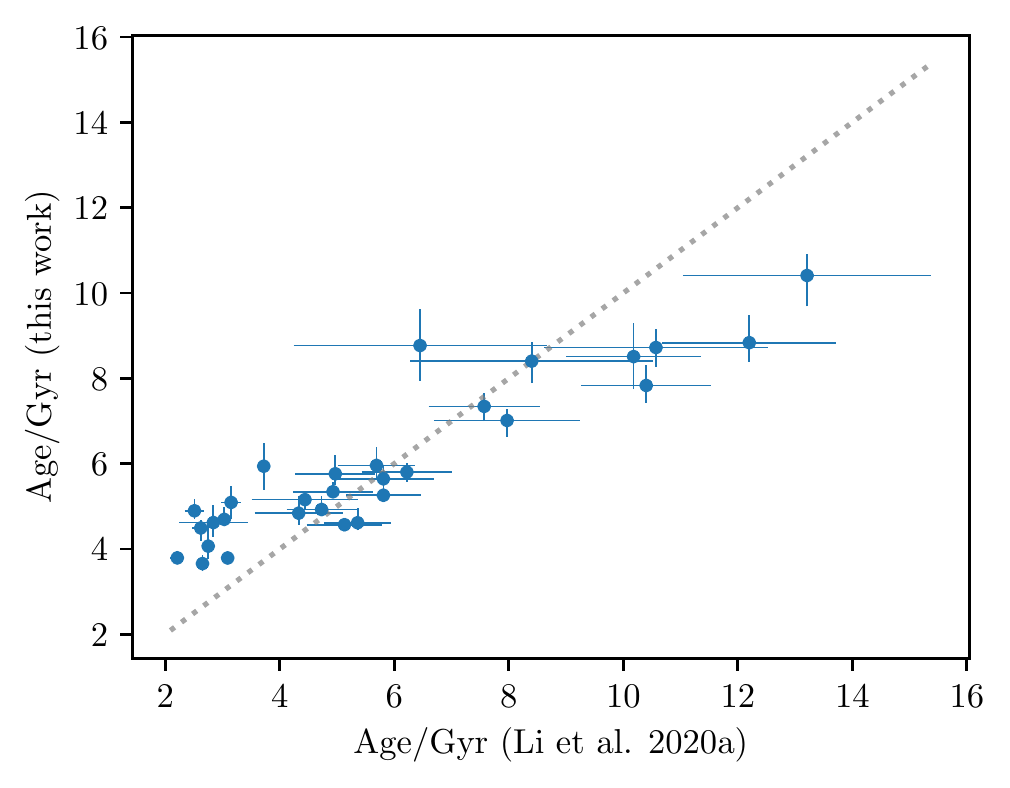}}{\node at (.9, .2){\textbf{(b)}};}
    \caption{Comparison of stellar masses (a) and ages (b) estimated from our grid vs. those from the model grid of \citet{li_ages_2020}, using identical prescriptions for the surface term (\bg\ to first order in mode coupling) and identical choices of spectroscopic constraints. \label{fig:li}}
\end{figure}

We show the results of these tests in the final column of
\cref{tab:pvalues}. Unfortunately, \citet{li_ages_2020} make no attempt
to infer the initial helium abundances of their sample; the grid used
for these estimates assumes an intrinsic metallicity-helium relation,
and so cannot meaningfully make such inferences. As such, we are unable
to perform a comparison of this kind for \(Y_0\). We find that the
differences in the inferred masses between our two pipelines are
statistically significant; we show a direct comparison of the two in
\cref{fig:li}a. Note that while the absolute systematic differences are
potentially larger than some of the effects we have seen in the previous
comparisons in \cref{fig:comp1,fig:comp2}, the errors reported by the
grid of \citet{li_ages_2020} are significantly larger than from our
grid, owing to their much coarser grid sampling and less complete
coverage of parameter space. Consequently, under the \(t\)-tests we have
conducted, which operate on quantities normalised by the reported
uncertainties, these differences are only significant at the \(3\sigma\)
level. This can also be seen in a similar comparison of the inferred
stellar ages (\cref{fig:li}b), where likewise large systematic
differences in the inferred ages are also seen, but are not significant
at the \(3\sigma\) level.

\hypertarget{discussion-and-conclusion}{%
\section{Discussion and Conclusion}\label{discussion-and-conclusion}}

\label{sec:discussion}

\citet{ong_differential_2021} showed that there exist qualitative
differences between how stellar properties estimated from grid-based
detailed seismic modelling respond to different methodological choices
in correcting for the surface term. On one hand, estimates of the
masses, radii, and ages of main-sequence stars appear largely
insensitive to the choice of surface term treatment, so long as one is
performed
\citep[cf.][]{basu_robustness_2018, compton_surface_2018, nsamba_asteroseismic_2018}.
Any observed normal modes in these stars are all p-modes. On the other
hand, estimates of these properties for a sample of red giants did
appear to depend significantly on this methodological choice. The
observed nonradial modes in this case are the most \(\pi\)-like of an
otherwise g-mode dominated eigenvalue spectrum. Neither of these results
offered any useful methodological guidance for the analysis of
subgiants, which lie in the intermediate regime where substantial
nonradial coupling occurs between the outer p-mode and interior g-mode
cavities. In order to extend that work to subgiants, we first had to
generalise existing surface-term treatments to correctly account for
mixed-mode coupling, which has hitherto been an open theoretical
problem. This was done in Paper I.

In this work, we have instead considered how these constructions affect
inferences of various stellar fundamental properties, for subgiants in
particular. We have found that unlike main-sequence stars, but like the
sample of red giants in \citet{ong_differential_2021}, the inferred
masses of stars in our subgiant sample do appear to depend
systematically on which treatment of the surface term is used in the
modelling procedure, and how mode coupling is handled in the process. A
similar result is seen to hold for the stellar ages, although not at
\(3\sigma\) significance. On the other hand, while significant
systematic differences in the inferred values of the initial helium
abundance \(Y_0\) still exist, they appear considerably less sensitive
to the choice of surface-term treatment for these subgiants than they
were for the main-sequence sample of \citet{ong_differential_2021}.
Finally, parameter values estimated using the generalised \bg~correction
appear to be somewhat more consistent with an additional luminosity
constraint than those from \(\epsilon\)-matching, once mode coupling is
fully accounted for. Taken together, these are suggestive of
intermediate behaviour between main-sequence and red-giant oscillators,
rather than there being a well-defined transition. However, the
comparisons here have not completely explored the differences between
these two extreme regimes. In particular, \citet{ong_differential_2021}
found a qualitative difference between properties inferred using
parametric surface-term corrections compared to nonparametric ones (via
both the \(\epsilon\)-matching algorithm and the method of separation
ratios) on the red giant branch, but not in main-sequence stars. A
similar comparison would require generalising the method of separation
ratios to mixed modes, for which no theoretical construction is
available at present.

As far as methodological guidance is concerned, we have found that fully
accounting for mode coupling in the surface term results in at least
different stellar masses compared to inferences made with first-order
approximations to mode coupling. Fortunately, the absolute effects of
changing the surface-term treatment appear to be still much smaller than
those of changing the underlying physics used in constructing stellar
models, or the sampling strategy used in grid searches. Nonetheless,
these surface-term methodological systematics are potentially
significant at the \(3\sigma\) level. Changing the prescription used for
the surface term also appears to induce systematic offsets in the
stellar age which are potentially comparable to, or larger than, those
resulting from e.g.~the introduction of other global constraints, like
the luminosity.

Based on this, we recommend that generalisations of a similar kind to
those we have described in Paper I be applied to prescriptions for
surface-term corrections in general, if they are to be used for
analysing mixed modes in evolved stars (especially with
\(\Dnu \lesssim 30\ \mu\)Hz). Since many existing parametric
surface-term corrections are either phenomenological or empirical in
nature
\citep[e.g.][]{kjeldsen_correcting_2008, sonoi_surface_2015, ball_surface_2017},
the precise form of such generalisations may not be obvious. On the
other hand, as described in Paper I, for more evolved stars these matrix
methods may not yet be fully practicable, both because of numerical
difficulties in evaluating the relevant coupling matrices, and
computational ones arising from the large rank of the matrices that are
obtained for evolved stars. In those cases, restriction to the
\(\pi\)-mode subsystem \citep[as done
in][]{ball_surface_2018, ong_differential_2021} might still ultimately
be necessary, at least at present.

Finally, we must further qualify that these results, both here and in
\citet{ong_differential_2021}, are meant to be representative of
multi-target grid searches, rather than single-target boutique
modelling. The multi-target strategy is typical of modelling pipelines
intended for use with large-scale surveys, such as those presently being
developed for the PLATO mission (cf.~Cunha et al.~in prep). It is more
difficult to make general statements about how different surface-term
treatments might affect the latter case, since such modelling work is by
construction performed ad hoc. To the extent that these generalised
treatments of the surface term may be more theoretically defensible for
mixed modes, however, it seems reasonable to us that these results
should also inform the decisions made in such boutique modelling.

\begin{acknowledgements}
The authors acknowledge the dedicated teams behind the \emph{Kepler} and K$2$ missions, without whom this work would not have been possible. Short-cadence data were obtained through the Cycle $1$-$6$ K$2$ Guest observer program. This work was partially supported by NASA K2 GO Award 80NSSC19K0102 and NASA TESS GO Award 80NSSC19K0374. We thank the Yale Center for Research Computing for guidance and use of the research computing infrastructure. Funding for the Stellar Astrophysics Centre is provided by The Danish National Research Foundation (Grant Agreement No.: DNRF106).

\software{NumPy \citep{numpy}, SciPy stack \citep{scipy}, AstroPy \citep{astropy:2013,astropy:2018}, Pandas \citep{pandas}, PBJam \citep{pbjam}, \diamonds\ \cite{diamonds}, \mesa\ \citep{mesa_paper_1,mesa_paper_2,mesa_paper_4}\footnote{\url{https://zenodo.org/communities/mesa/?page=1&size=20}}, \gyre\ \citep{townsend_gyre_2013}.}
\end{acknowledgements}

\appendix

We provide the mode frequencies of all the stars in our K2 sample
(including the main-sequence and red giant stars not explicitly modelled
in this work) in \cref{tab:k2seis}.

\begin{longtable*}[]{@{}rccl@{}}
\caption{Mode frequencies for stars in K2 sample. A full list of mode
frequencies is available as a machine-readable
table.\label{tab:k2seis}}\tabularnewline
\toprule
EPIC & \(l\) & \(\nu/\mu\)Hz & \(\sigma_\nu/\mu\)Hz \\
\midrule
\endfirsthead
\toprule
EPIC & \(l\) & \(\nu/\mu\)Hz & \(\sigma_\nu/\mu\)Hz \\
\midrule
\endhead
212478598 & 0 & 472.498571 & 0.009596 \\
212478598 & 0 & 507.598629 & 0.019181 \\
212478598 & 0 & 542.358328 & 0.008893 \\
212478598 & 0 & 577.638506 & 0.012759 \\
212478598 & 0 & 612.959501 & 0.012075 \\
212478598 & 1 & 421.465770 & 0.010655 \\
212478598 & 1 & 451.967692 & 0.008527 \\
212478598 & 1 & 462.127731 & 0.012090 \\
212478598 & 1 & 482.719768 & 0.035494 \\
212478598 & 1 & 493.325932 & 0.012059 \\
\(\vdots\) & & & \\
\bottomrule
\end{longtable*}

  \bibliography{biblio.bib}

\clearpage

\begin{figure*}[p]
    \centering\vspace*{-10ex}
    \annotate{\includegraphics[width=.425\textwidth, trim=.25cm .25cm .25cm .15cm,clip]{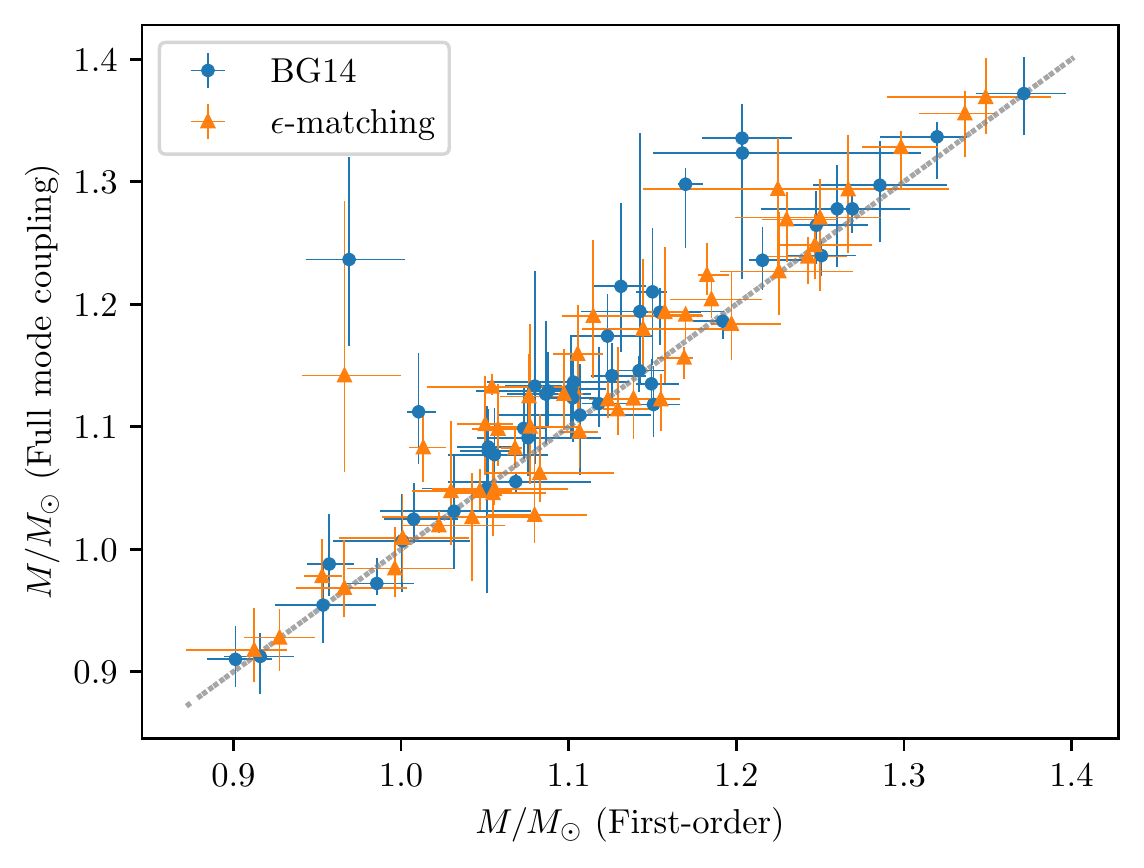}}{\node at (.93, .2){\textbf{(a)}};}
    \annotate{\includegraphics[width=.425\textwidth, trim=.25cm .25cm .25cm .15cm,clip]{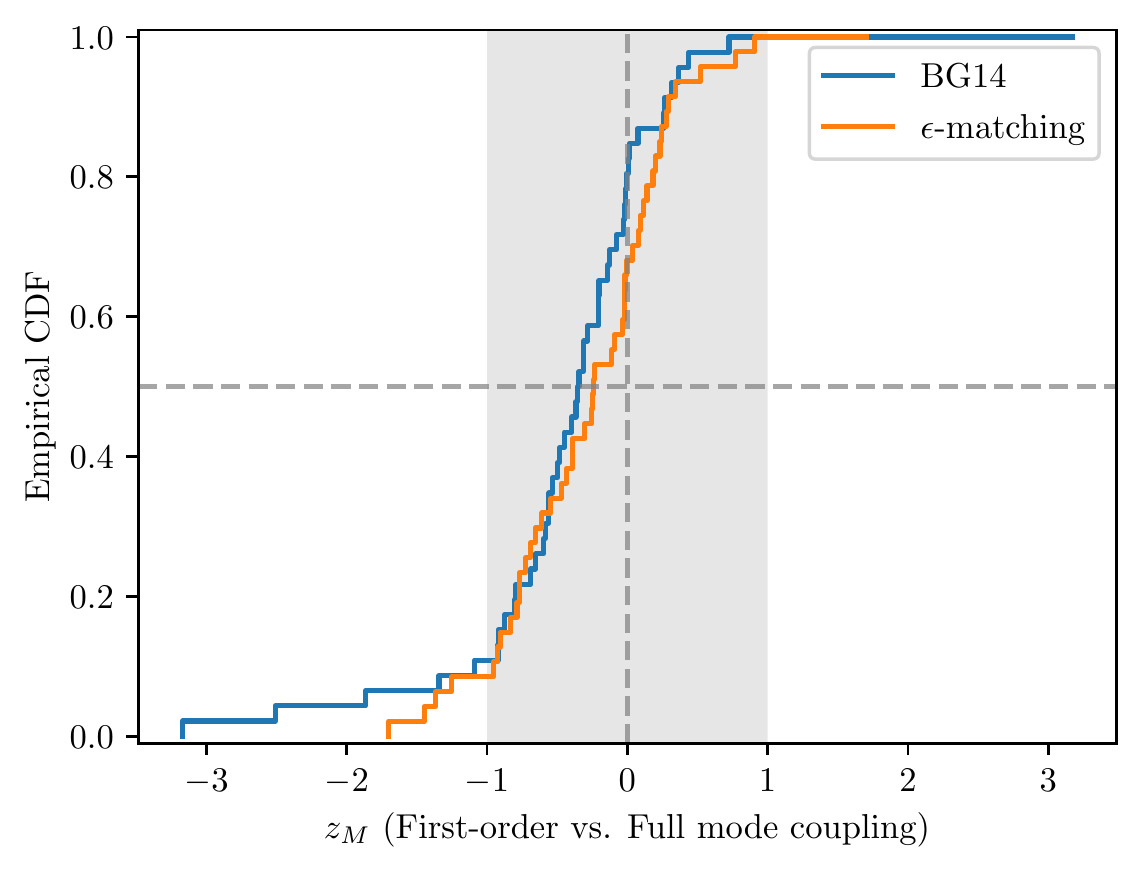}}{\node at (.93, .2){\textbf{(e)}};}
    \annotate{\includegraphics[width=.425\textwidth, trim=.25cm .25cm .25cm .15cm,clip]{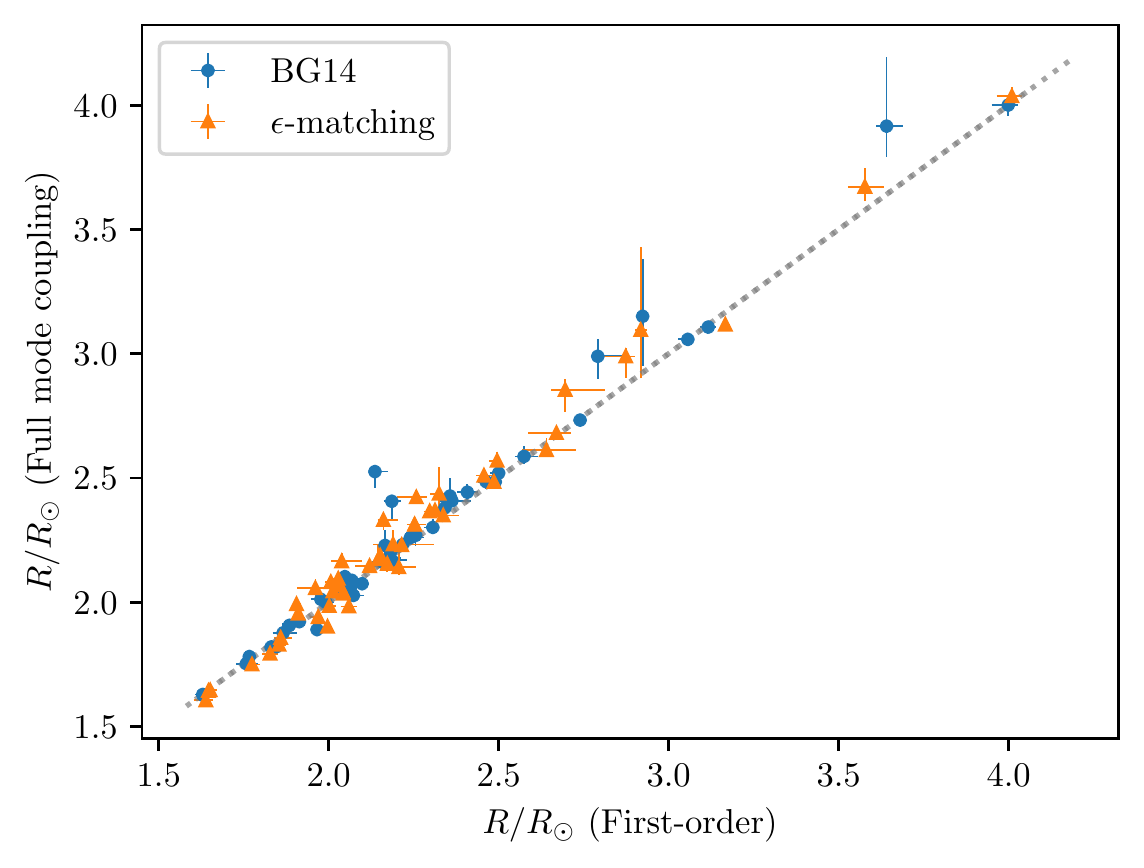}}{\node at (.93, .2){\textbf{(b)}};}
    \annotate{\includegraphics[width=.425\textwidth, trim=.25cm .25cm .25cm .15cm,clip]{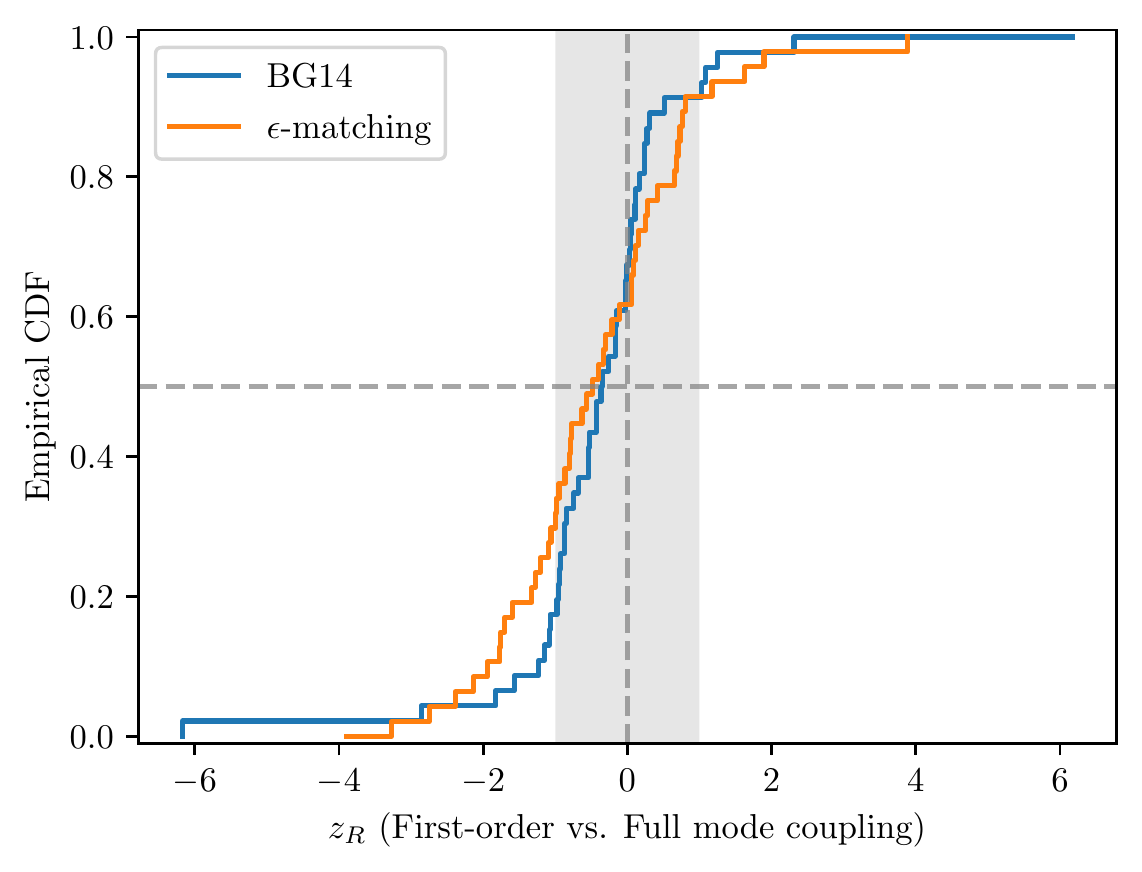}}{\node at (.93, .2){\textbf{(f)}};}
    \annotate{\includegraphics[width=.425\textwidth, trim=.25cm .25cm .25cm .15cm,clip]{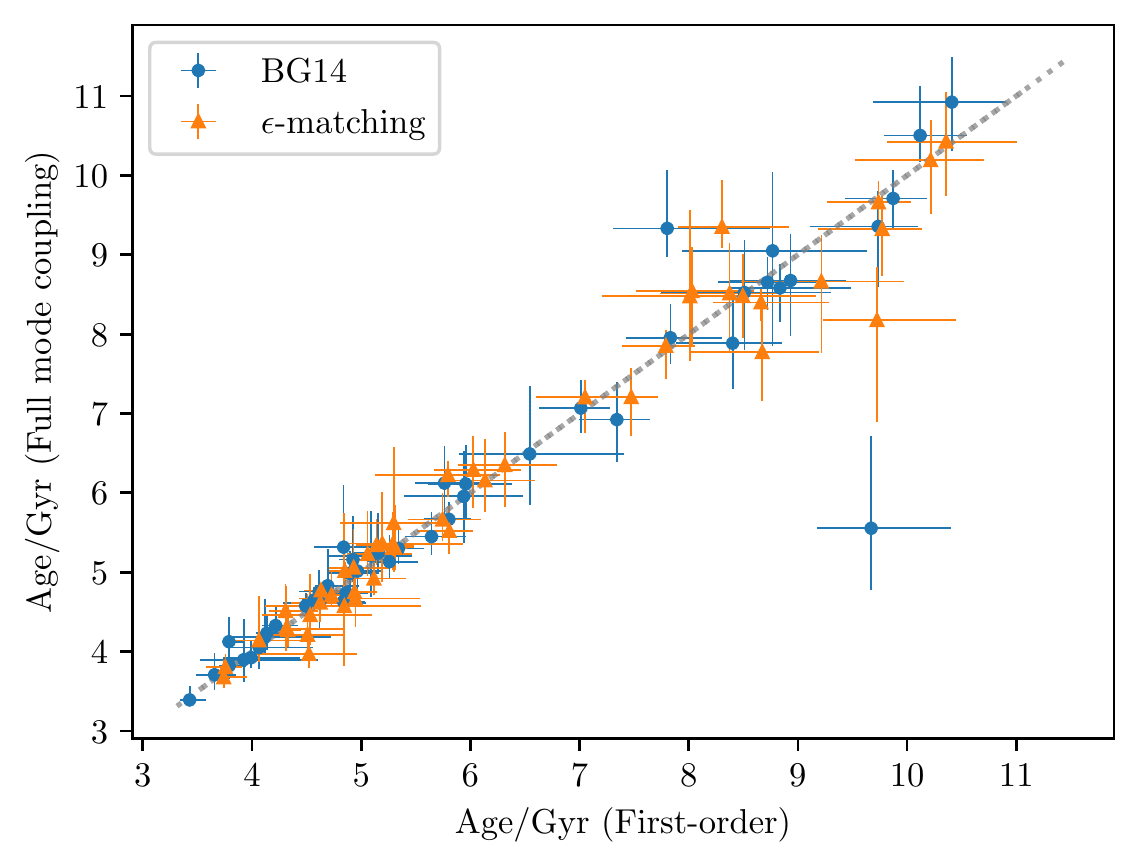}}{\node at (.93, .2){\textbf{(c)}};}
    \annotate{\includegraphics[width=.425\textwidth, trim=.25cm .25cm .25cm .15cm,clip]{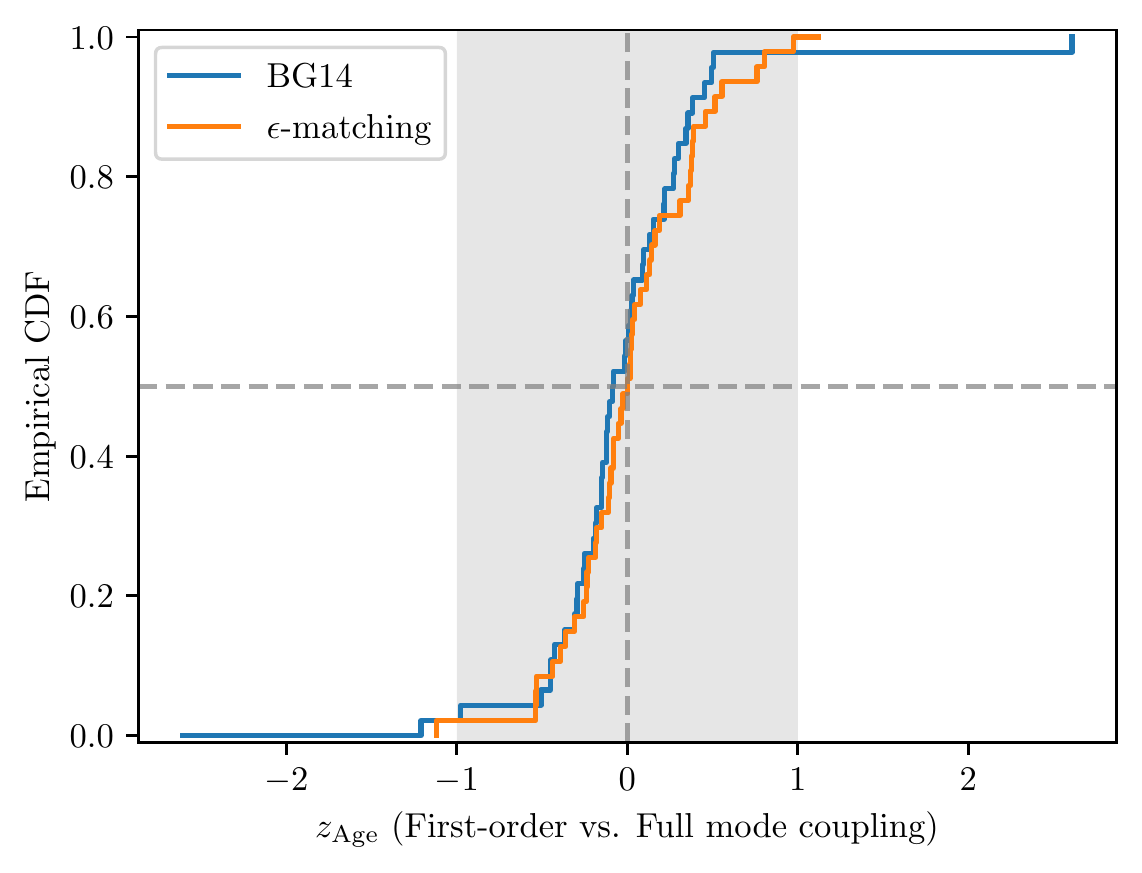}}{\node at (.93, .2){\textbf{(g)}};}
    \annotate{\includegraphics[width=.425\textwidth, trim=.25cm .25cm .25cm .15cm,clip]{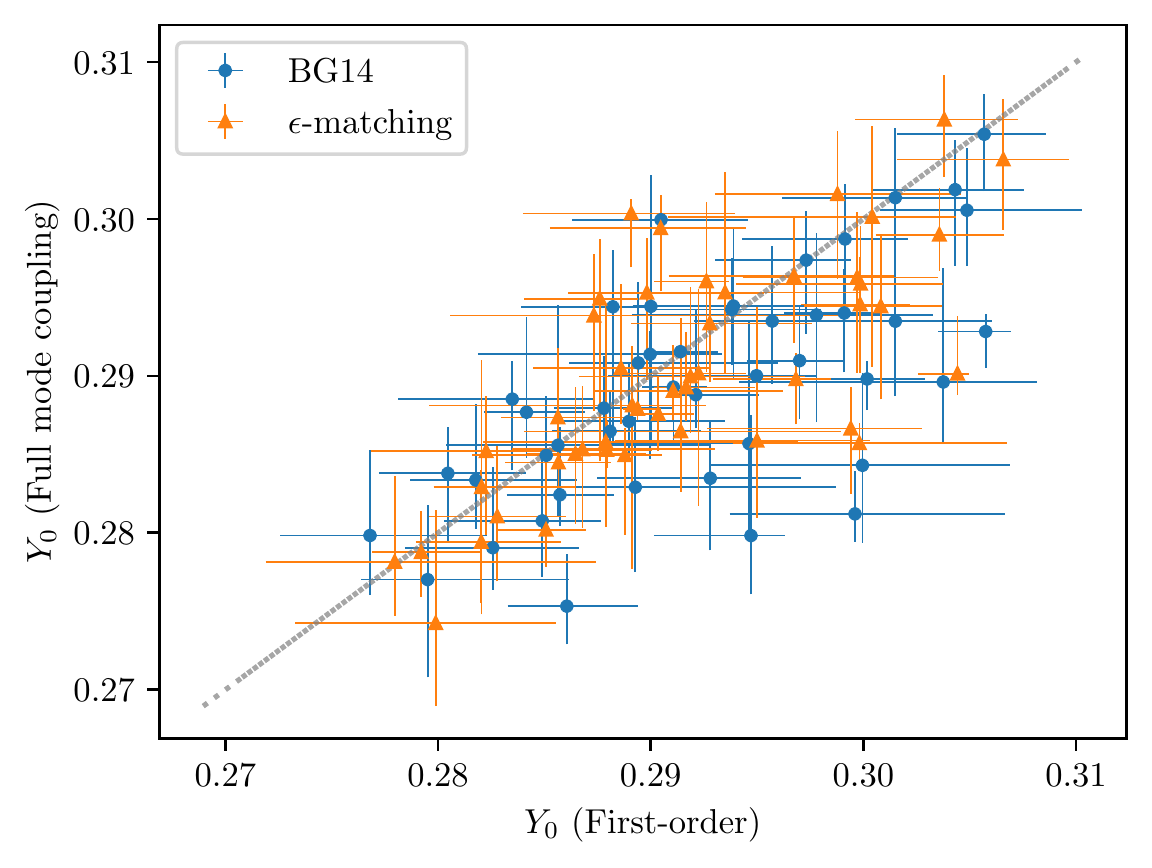}}{\node at (.93, .2){\textbf{(d)}};}
    \annotate{\includegraphics[width=.425\textwidth, trim=.25cm .25cm .25cm .15cm,clip]{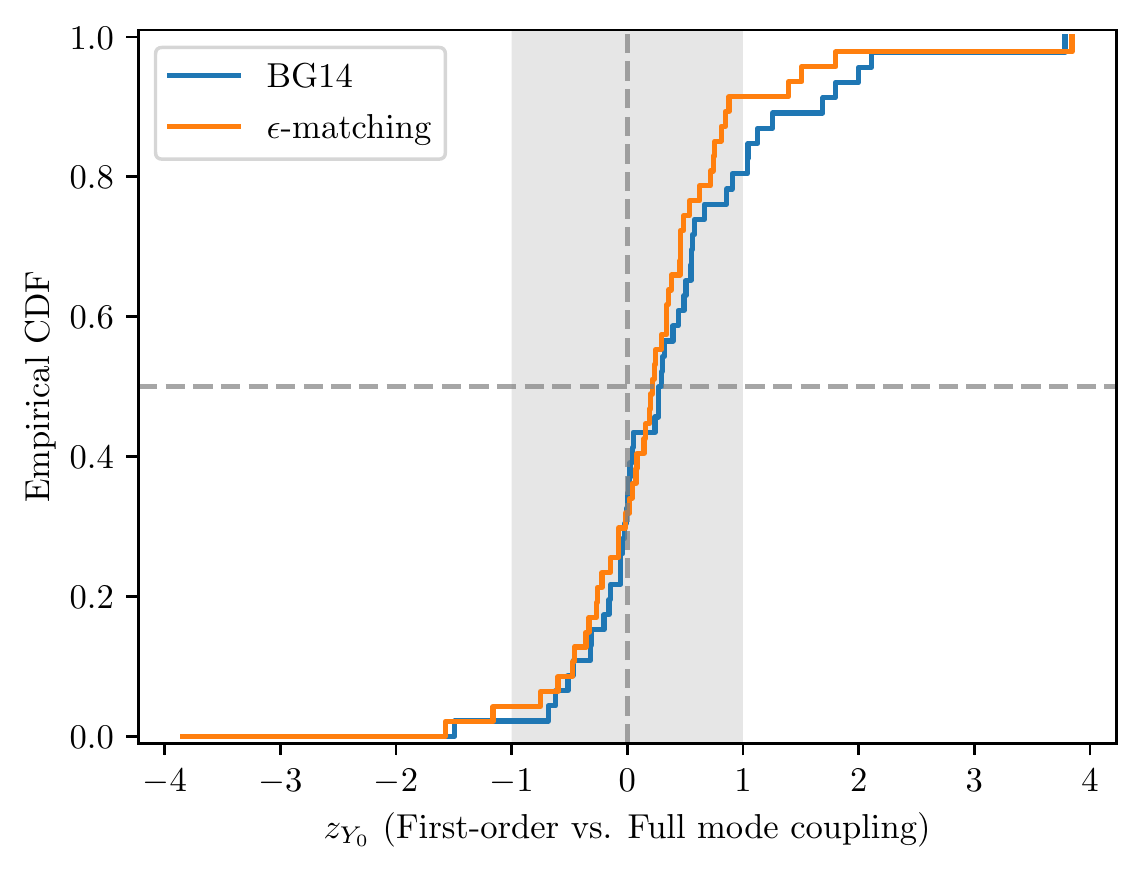}}{\node at (.93, .2){\textbf{(h)}};}
    \caption{Comparisons between properties estimated using first-order and full treatments of mode coupling for two different surface-term corrections. In Figs. \ref{fig:comp1}a-\ref{fig:comp1}d (left column) we show a direct comparison between first-order (horizontal axis) and full (vertical) mode coupling calculations, with points showing the median of the posterior distribution, coloured by theoretical prescription. The line of equality is shown with the grey dotted line in the background. In Figs. \ref{fig:comp1}e-\ref{fig:comp1}h (right column) we show the empirical cumulative distributions of the normalised differences, per \cref{eq:z}. We show the $\pm 1\sigma$ interval with the shaded region. Dashed lines indicate $z=0$ and the theoretical median ($p=0.5$) of the CDF.}
    \label{fig:comp1}
\end{figure*}

\begin{figure*}[p]
    \centering\vspace*{-1ex}
    \annotate{\includegraphics[width=.425\textwidth, trim=.25cm .25cm .25cm .15cm,clip]{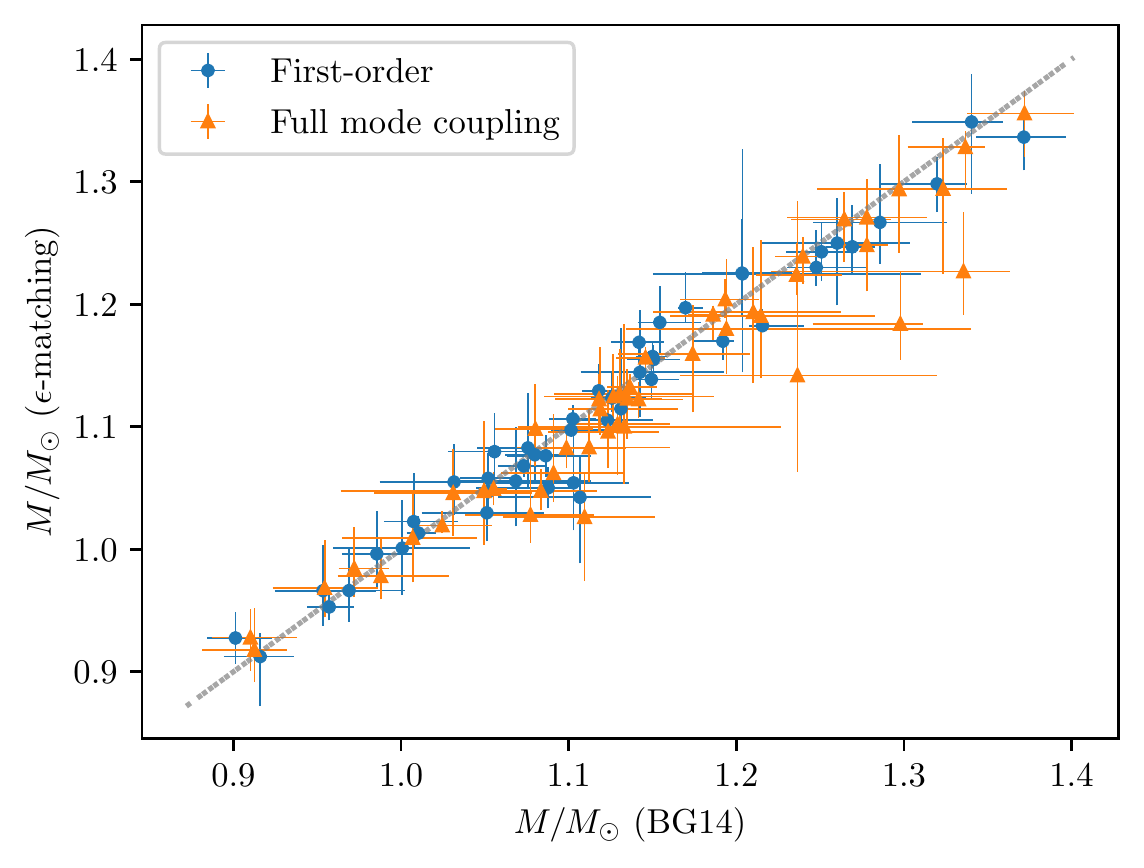}}{\node at (.93, .2){\textbf{(a)}};}
    \annotate{\includegraphics[width=.425\textwidth, trim=.25cm .25cm .25cm .15cm,clip]{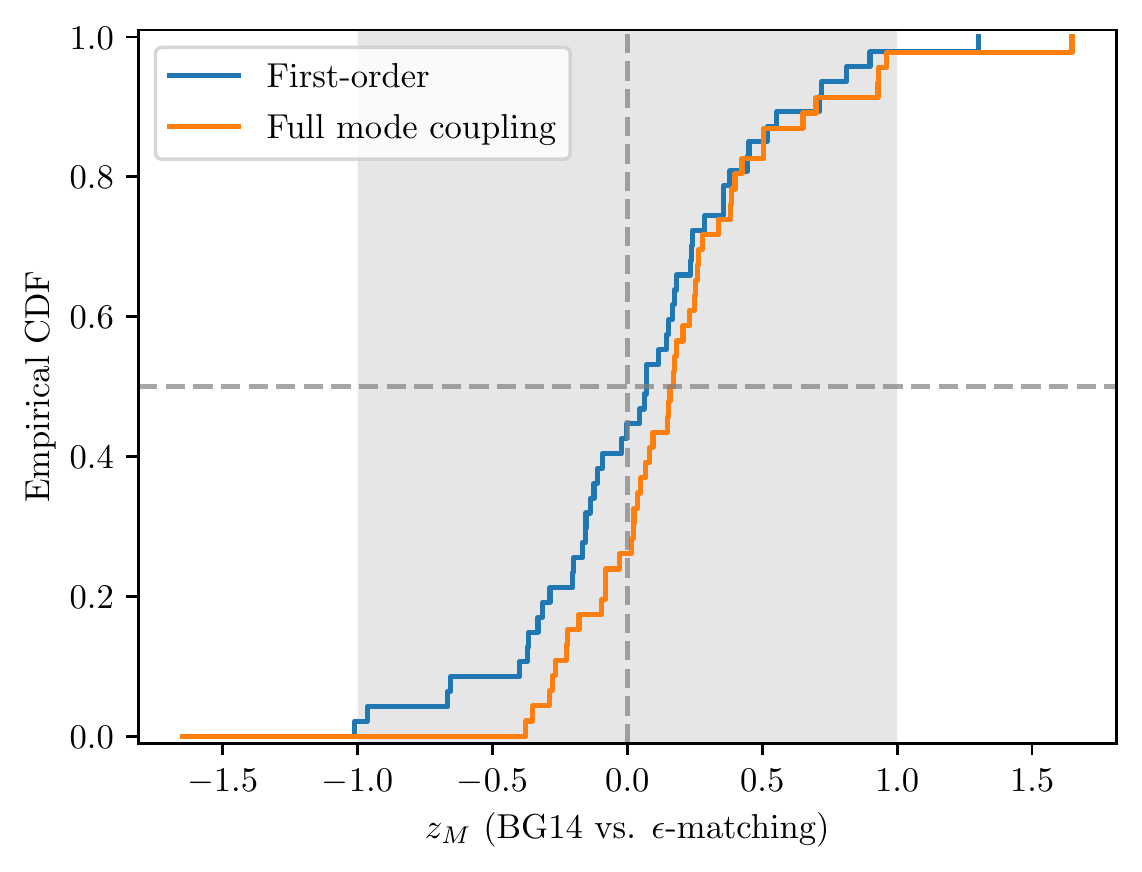}}{\node at (.93, .2){\textbf{(e)}};}
    \annotate{\includegraphics[width=.425\textwidth, trim=.25cm .25cm .25cm .15cm,clip]{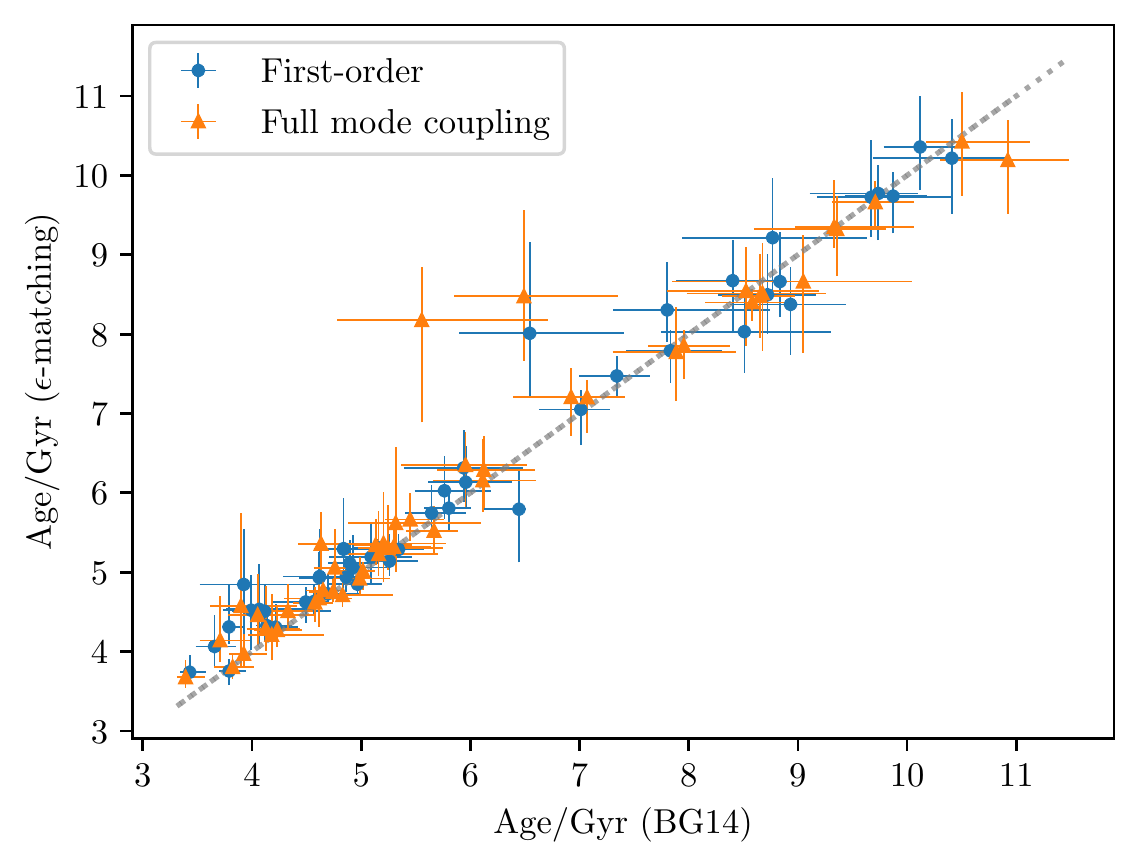}}{\node at (.93, .2){\textbf{(b)}};}
    \annotate{\includegraphics[width=.425\textwidth, trim=.25cm .25cm .25cm .15cm,clip]{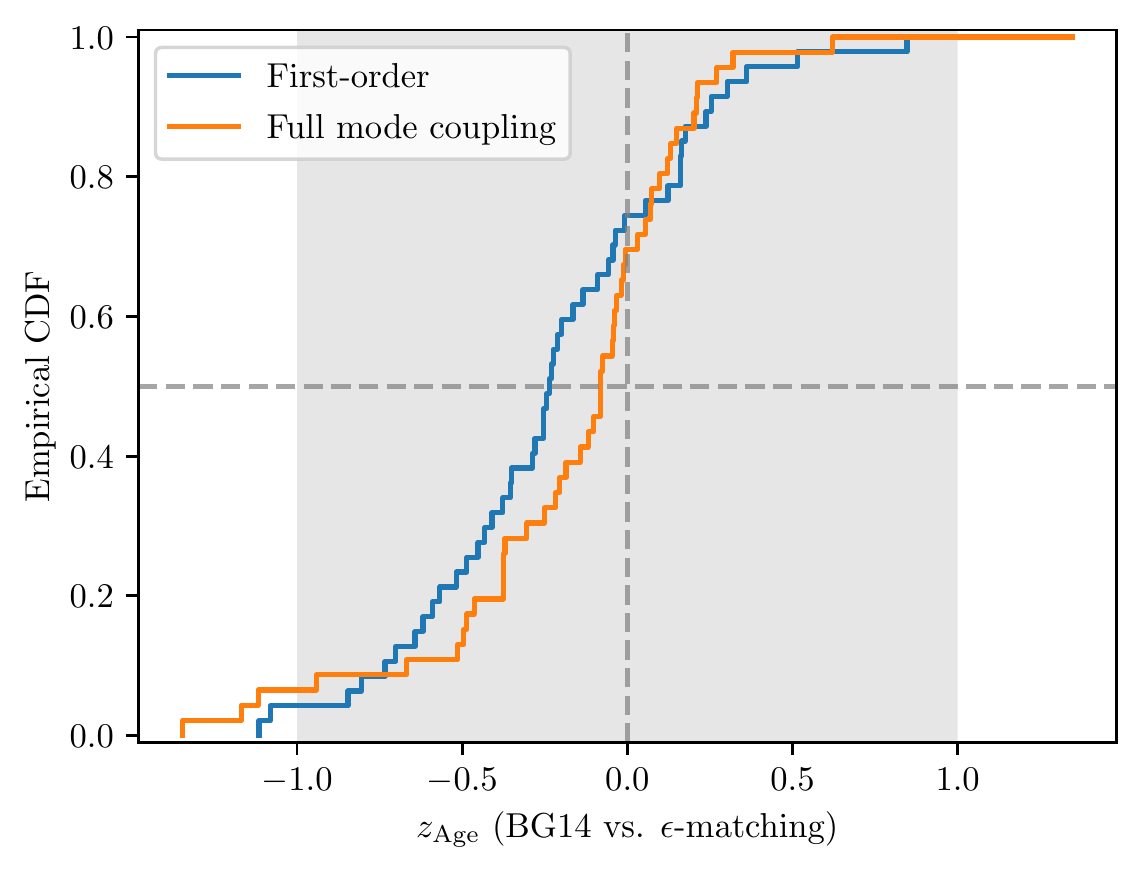}}{\node at (.93, .2){\textbf{(f)}};}
    \annotate{\includegraphics[width=.425\textwidth, trim=.25cm .25cm .25cm .15cm,clip]{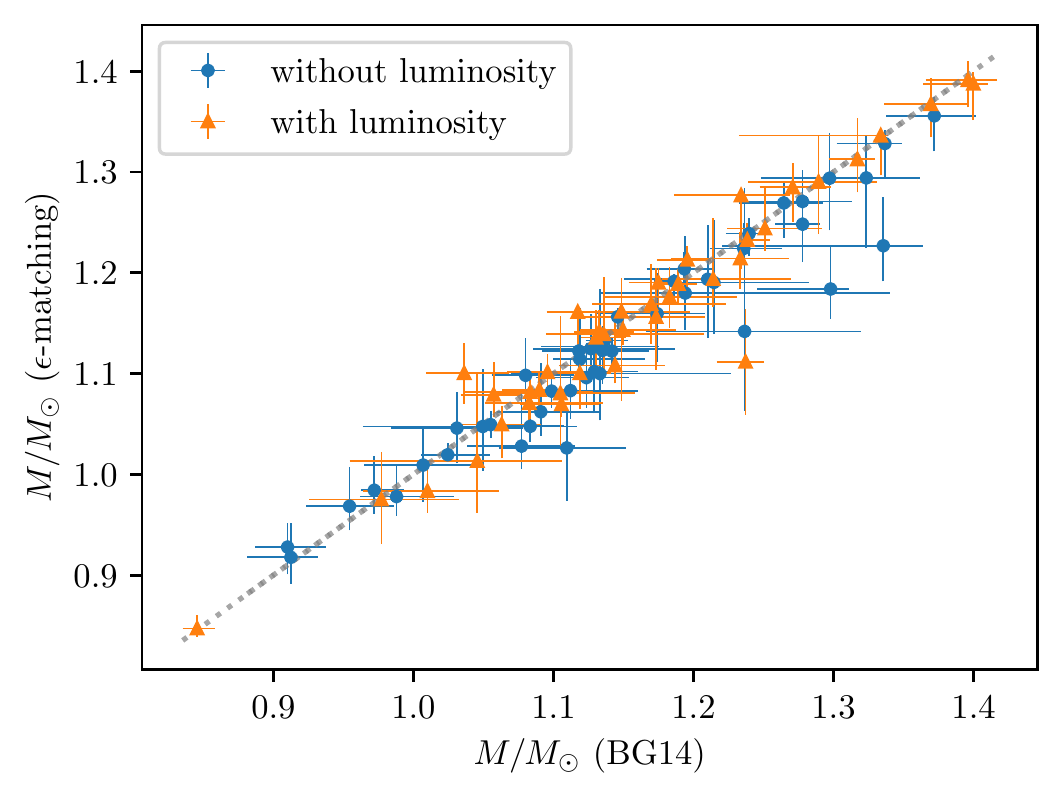}}{\node at (.93, .2){\textbf{(c)}};}
    \annotate{\includegraphics[width=.425\textwidth, trim=.25cm .25cm .25cm .15cm,clip]{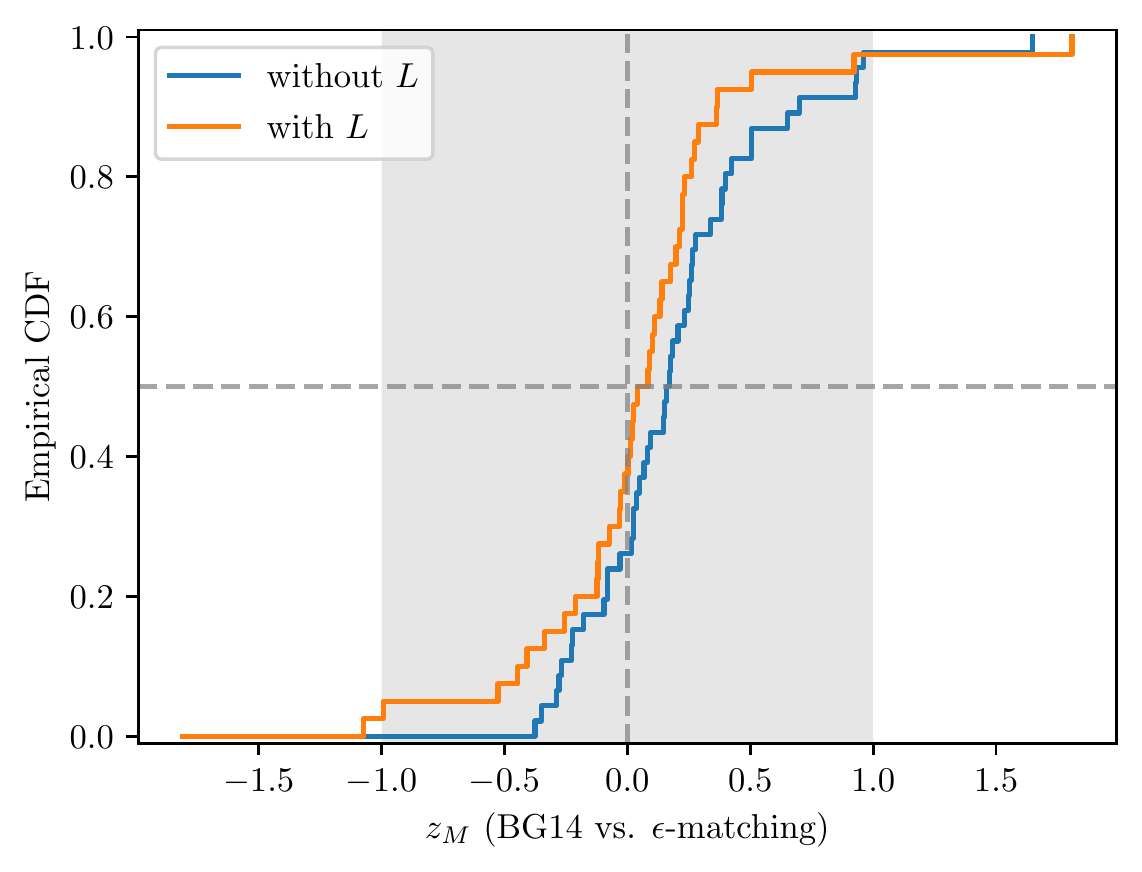}}{\node at (.93, .2){\textbf{(g)}};}
    \annotate{\includegraphics[width=.425\textwidth, trim=.25cm .25cm .25cm .15cm,clip]{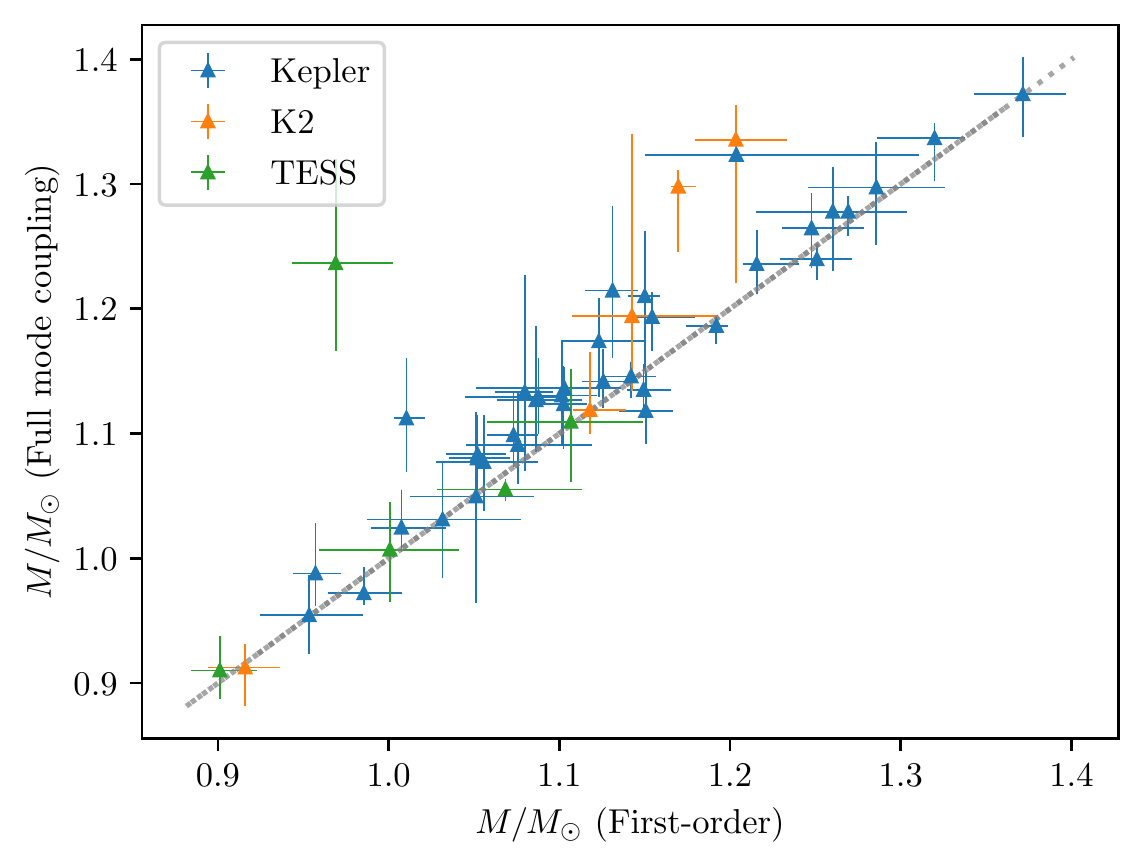}}{\node at (.93, .2){\textbf{(d)}};}
    \annotate{\includegraphics[width=.425\textwidth, trim=.25cm .25cm .25cm .15cm,clip]{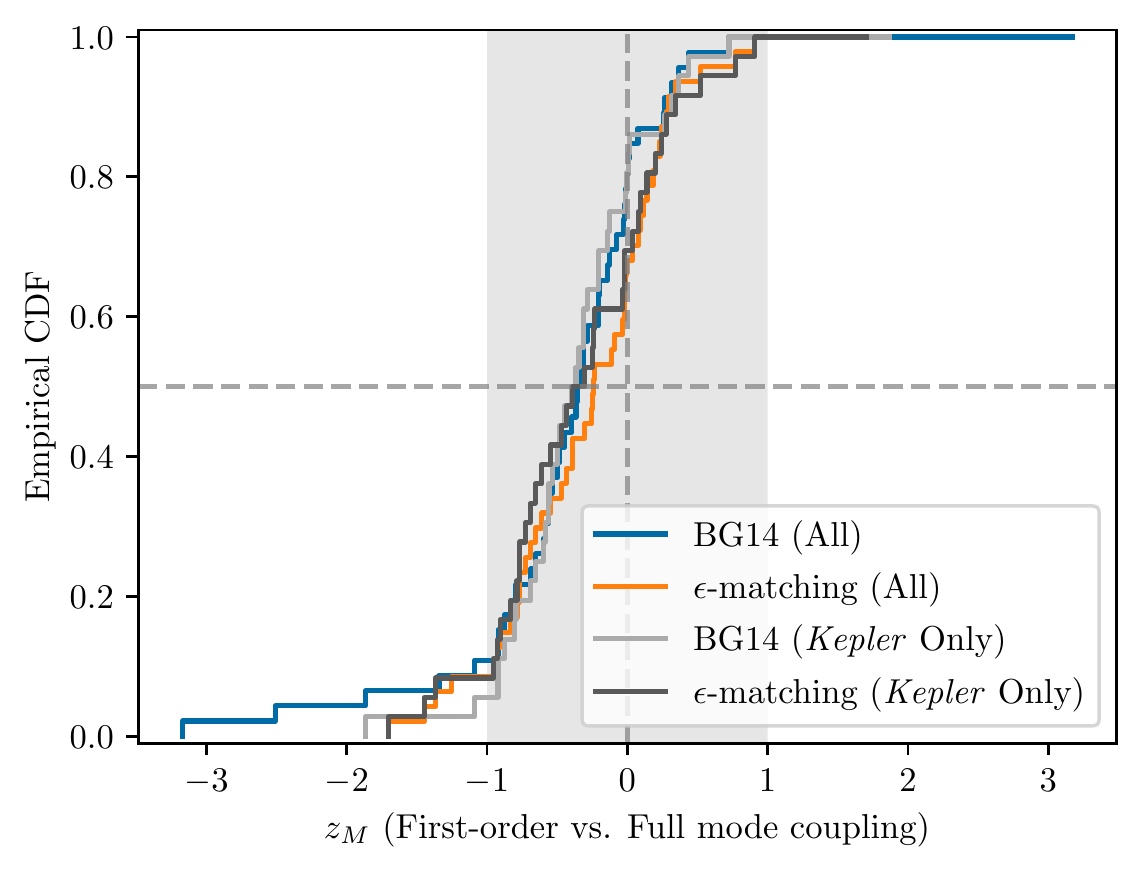}}{\node at (.15, .93){\textbf{(h)}};}
    \caption{Various pairwise comparisons, as in \cref{fig:comp1}. In Figs. \ref{fig:comp2}a-\ref{fig:comp2}d (left column) we show direct comparison between inferred quantities, and in Figs. \ref{fig:comp2}e-\ref{fig:comp2}h (right column) we show the empirical cumulative distributions of the normalised differences, per \cref{eq:z}.}
    \label{fig:comp2}
\end{figure*}

\end{document}

%% file: macros.tex
\usepackage[capitalize]{cleveref}
\usepackage{CJKutf8}

\newcommand{\Dnu}{\ensuremath{\Delta\nu}}
\newcommand{\numax}{\ensuremath{{\nu_\text{max}}}}
\newcommand{\amlt}{\ensuremath{{\alpha_\text{MLT}}}}
\newcommand{\fov}{\ensuremath{{f_\text{ov}}}}
\newcommand{\teff}{\ensuremath{{T_\text{eff}}}}
\newcommand{\logg}{\ensuremath{{\log g}}}
\newcommand{\feh}{\ensuremath{{[\text{Fe/H}]}}}
\newcommand{\chinesename}{{\begin{CJK}{UTF8}{gbsn}(王加冕)\end{CJK}}}

\newcommand{\annotate}[2]{\begin{tikzpicture}
    \node[anchor=south west,inner sep=0,align=center] (image) at (0,0) {
    #1
    };
    \begin{scope}[x={(image.south east)},y={(image.north west)}]
    #2
    \end{scope}
\end{tikzpicture}}

\def\bibitem{\vskip\bibskip\footnotesize\savebibitem}

%% file: preamble.tex
\correspondingauthor{Joel Ong}
\email{joel.ong@yale.edu}
\author[0000-0001-7664-648X]{J. M. Joel Ong \chinesename}
\affiliation{Department of Astronomy, Yale University, 52 Hillhouse Ave., New Haven, CT 06511, USA}

\author[0000-0002-6163-3472]{Sarbani Basu}
\affiliation{Department of Astronomy, Yale University, 52 Hillhouse Ave., New Haven, CT 06511, USA}

\author[0000-0001-9214-5642]{Mikkel N. Lund}
\affil{Stellar Astrophysics Centre, Department of Physics and Astronomy, Aarhus University, Ny Munkegade 120, DK-8000 Aarhus C, Denmark}

\author[0000-0001-6637-5401]{Allyson Bieryla}
\affiliation{Center for Astrophysics, Harvard \& Smithsonian, 60 Garden Street, Cambridge, MA 02138, USA}

\author[0000-0002-7541-9346]{Lucas S. Viani}
\affiliation{Department of Astronomy, Yale University, 52 Hillhouse Ave., New Haven, CT 06511, USA}

\author[0000-0001-9911-7388]{David W. Latham}
\affiliation{Center for Astrophysics, Harvard \& Smithsonian, 60 Garden Street, Cambridge, MA 02138, USA}

\received{June 4, 2021}
\revised{July 23, 2021}
\accepted{August 16, 2021}
\submitjournal{\apj}

